\documentclass[sn-mathphys-num,iicol]{sn-jnl}

\usepackage{graphicx}%
\usepackage{multirow}%
\usepackage{amsmath,amssymb,amsfonts}%
\usepackage{amsthm}%
\usepackage{mathrsfs}%
\usepackage{mathtools}
\usepackage[title]{appendix}%
\usepackage{xcolor}%
\usepackage{textcomp}%
\usepackage{manyfoot}%
\usepackage{booktabs}%
\usepackage{algorithm}%
\usepackage{algorithmicx}%
\usepackage{algpseudocode}%
\usepackage{dblfloatfix}
\usepackage{listings}%
\usepackage{physics}
\usepackage{cancel}
\usepackage{multicol}
\usepackage{dcolumn}
\usepackage{bm}
\usepackage{graphicx,caption,subcaption}
\usepackage{url}
\usepackage[normalem]{ulem}
\usepackage{enumitem}
\usepackage{bbding}
\usepackage{pifont}
\usepackage{latexsym, array,multirow,  verbatim, enumerate}
\usepackage{natbib}

\begin{document}

\title[Article Title]{Resurgence in the Scalar Quantum Electrodynamics Euler-Heisenberg Lagrangian}


\author[1,2]{\fnm{Drishti} \sur{Gupta}}\email{dg36@illinois.edu}

\author[1]{\fnm{Arun M.} \sur{Thalapillil}}\email{thalapillil@iiserpune.ac.in}

\affil[1]{\orgdiv{Department of Physics}, \orgname{Indian Institute of Science Education and Research Pune}, \orgaddress{\street{Homi Bhabha Road, Pashan}, \city{Pune}, \postcode{411008}, \country{Pune}}}

\affil[2]{\orgdiv{Department of Physics}, \orgname{University of Illinois Urbana-Champaign}, \orgaddress{\city{Urbana}, \state{IL}, \postcode{61801}}}


\abstract{
We explore the ideas of resurgence and Pad\'{e}-Borel resummation in the Euler-Heisenberg Lagrangian of scalar quantum electrodynamics, which has remained largely unexamined in these contexts. We thereby extend the related seminal works in spinor quantum electrodynamics, while contrasting the similarities and differences in the two cases. We investigate in detail the efficacy of resurgent extrapolations starting from just a finite number of terms in the weak-field expansions of the 1-loop and 2-loop scalar quantum electrodynamics Euler-Heisenberg Lagrangian. While we re-derive some of the well-known 1-loop and 2-loop contributions in representations suitable for Pad\'{e}-Borel analyses, other contributions have been derived for the first time. For instance, we find a closed analytic form for the one-particle reducible contribution at 2-loop, which until recently was thought to be zero. It is pointed out that there could be an interesting interplay between the one-particle irreducible and one-particle reducible terms in the strong-field limit. The 1-loop scalar electrodynamics contribution may be effectively mapped into two copies of the spinor quantum electrodynamics, and the particle reducible contribution may be mapped to the 1-loop contribution. It is suggested that these mappings cannot be trivially used to map the corresponding resurgent structures. The singularity structures in the Pad\'{e}-Borel transforms at 1-loop and 2-loop are examined in some detail. Analytic continuation to the electric field case and the generation of an imaginary part is also studied. We compare the Pad\'{e}-Borel reconstructions to closed analytic forms or to numerically computed values in the full theory.}
\maketitle
\section{Introduction}

Understanding the diverse facets of quantum field theories and their extensions is an ongoing endeavour (See, for instance,\,\cite{Polyakov:1987ez,Deligne:1999qp,Witten:1997fz,Witten:2018zxz} and related references). These aspirations include both gaining a better understanding of the various mathematical structures that arise in the variegated manifestations of quantum field theories as well as developing effective methodologies to apply appropriate quantum field theories to physical scenarios of interest. 

The calculation of quantities in interacting theories is often a non-trivial task. In many such situations, one resorts to perturbative methods. These perturbative series for an observable of interest ($\mathcal{O}$) involving a small parameter ($\epsilon \rightarrow 0$)---for instance, a weak coupling constant, low temperature, or small field value---are in general a power series in the small parameter
\begin{equation}
\mathcal{O}(\epsilon)\overset{?}=\sum_{k\geq0}\,c^{(0)}_k \, \epsilon^k \; .
\label{eq:pertseries}
\end{equation}
Interestingly, these perturbative series are more often than not asymptotic series with vanishing radii of convergence (for a nice pedagogical discussion of some of these aspects, please see, for instance,\,\cite{Bender1999,marino_2015}). Dyson was among the first to realize that, in many cases, the lack of convergence in a perturbative series is related to the presence of instabilities in other sectors of the theory\,\cite{dyson}. 

A question then arises about correctly interpreting the oft-written equality in eq.\,(\ref{eq:pertseries}) and broadly the perturbative series itself. In this context, a related consideration is if there is a way to incorporate more information from the asymptotic series towards the reconstruction of the full observable $\mathcal{O}$ than the usual optimal truncation of the said asymptotic series.

Resummation methods and, more broadly, the mathematical frameworks of resurgence theory and trans-series\,\cite{ecalle1985fonctions,10.1093/oso/9780198535850.002.0003, Sternin:1996,Costin:2008} address these pertinent questions, and attempt to reconstruct the complete function from just the perturbative expansion (see also, for instance, \cite{DORIGONI2019167914,Aniceto:2018bis}). The observable $\mathcal{O}$ may be extended to incorporate non-perturbative aspects, occasionally with ambiguities, from the associated small-parameter perturbative series by applying resurgent analysis and trans-series constructions. The latter is a natural extension of the perturbative series in eq.\,(\ref{eq:pertseries}) around $\epsilon=0$, and in its simplest construction may be of a form\,\cite{ecalle1985fonctions,DORIGONI2019167914,Aniceto:2018bis}
\begin{equation}
\mathcal{O}(\epsilon) =\sum_{k\geq0}\,c^{(0)}_k \, \epsilon^k+ \sum_{I\geq 1} \exp\left[-\frac{C_I}{\epsilon}\right] \sum_{l\geq0}\,c^{(I)}_l \, \epsilon^l \; .
\label{eq:transs}
\end{equation}

Terms of the form $\exp\left[-C_I/\epsilon\right]$ have an essential singularity at $\epsilon=0$ and are invisible at any order in the small-parameter perturbative series. If the original perturbative series is interpreted as an expansion about the free theory, the other terms are heuristically like an expansion about the other non-trivial saddle points that may be present. The framework of resurgence and Borel resummation provides a method to connect the various parts of this trans-series---for instance, relating the large-order behavior of $c_k^{(0)}$ to $\exp\left[-C_1/\epsilon\right]$. Remarkably, even when only a finite number of terms are accessible to us in the perturbative series, one may, in many cases, via the method of Pad\'{e} approximants (see\,\cite{baker1975essentials,baker1996pade}, for instance), perform resummations to yield accurate reconstructions of the full function. 

Since its inception, resurgent analysis has been applied to various cases---for example, see\,\cite{Voros1983, Jentschura:2004jg, Dunne:2012ae, Argyres:2012ka, Dunne:2012zk, Marino:2012zq,Aniceto:2014hoa,Dunne:2016nmc}. The study of resurgence has continued to be a very active area of research as well in the past few years, for instance, to give a random sampling of recent studies---in the study of the large-order behaviour in quantum chromodynamics\,\cite{MiravitllasMas:2019neb}, investigating resurgent extrapolation with the Painleve I equation as a prototype\,\cite{Costin:2019xql}, study of superconductivity\,\cite{Marino:2019wra}, the study of renormalisation group equations\,\cite{Bersini:2019axn}, applications to the one-dimensional Hubbard model\,\cite{Marino:2020dgc}, study of quantum knot invariants\,\cite{Garoufalidis:2020nut}, the study of renormalons in six-dimensional $\phi^3$ theories\,\cite{Borinsky:2022knn}, investigations of the $O(3)$ and $O(4)$ sigma models\,\cite{Bajnok:2021zjm,Bajnok:2021dri}, analysis of integrable field theories\,\cite{DiPietro:2021yxb}, large-charge expansions in $O(2N)$ models\,\cite{Dondi:2021buw}, analysis of $SU(2)$ Chern-Simons partition function\,\cite{Wu:2020dhl}, study of $sl(2,\mathbb{C})$ Chern-Simons state integral models\,\cite{Duan:2022ryd}, the explicit construction of four-dimensional de Sitter space via nodal diagrams emerging from Glauber-Sudarshan states\,\cite{Brahma:2022wdl}, in the context of matrix-models and D-branes\,\cite{Schiappa:2023ned}, in the context of conformal blocks and the large central-charge limit\,\cite{Benjamin:2023uib}, study of supersymmetric $\mathcal{N}=1$  Jackiw-Teitelboim gravity \,\cite{Griguolo:2023jyy}, studies at the intersection of quantum topology and mathematical physics\,\cite{Costin:2023kla}, to mention a few.

On the other hand, the study of quantum field theories in the strong-field regime has an extensive history and is of much theoretical and experimental current interest (for recent discussions, see, for instance,\,\cite{Altarelli:2019zea, Hattori:2023egw, Fedotov_2023} and related references). Understanding strong-field effects is also of relevance in astrophysical and cosmological contexts\,\cite{Ruffini:2009hg,Kaspi:2017fwg,Kim:2019joy,Grasso:2000wj,Durrer:2013pga}, where such terrestrially-inaccessible fields and conditions may abound.

Apart from the theoretical and observational motivations, recent advancements in generating ultra-intense background fields terrestrially have sparked renewed efforts in the study of the applications of Quantum Electrodynamics (QED) in the strong-field regime\,\cite{Fedotov_2023}. The E-144 experiment at SLAC was the first of its kind to observe non-linear scattering of light in intense background fields\,\cite{SLAC-E144-Non-Linear-QED,SLAC-E144-Compton-Scattering,SLAC-E144-Positron-Production}. Since then, multiple experimental setups such as ELI-NP\,\cite{ELI-NP} and LUXE\,\cite{LUXE} have been proposed to achieve even higher intensities of laser fields. The physics at ultra-high intensities of background fields, in addition to non-linear scattering effects, is also influenced by non-perturbative effects that cannot be calculated in the paradigm of usual perturbation theory. One such effect is ``Schwinger pair production", which is the decay of the QED vacuum to produce pairs of particle and anti-particle in the presence of intense electric fields\,\cite{Sauter:1931zz,Heisenberg:1936nmg,Schwinger_1951}. The Schwinger pair-production rate is exponentially suppressed by a factor of $e^{-\pi E_c/E}$, where $E$ is the electric field, $E_c = m^2/q$ (we adopt natural units, with $c= \hbar = 1$, in the study) is the critical value of the electric field, $m$ is the mass of the pair-produced particles and $q$ are their charges. There are a few open and interesting problems in strong-field (spinor) QED and scalar QED (sQED) that have come to the fore in recent years (for example, see \,\cite{Ruffini:2009hg,Medina:2015qzc,Draper:2018lyw,Gould:2017fve,Gould:2018ovk,Korwar:2018euc,Torgrimsson:2019sjn,Gould:2018efv}). In many of these studies the world line instanton framework has proven to be a powerful tool\,\cite{Bern:1991aq, Strassler:1992zr}. The effective action in this scheme is represented as a one-dimensional path integral, and the stationary worldline loops which appear in this context are called worldline instantons\,\cite{Affleck:1981bma, Dunne:2005sx, Dunne:2006st}. These instantons will be of relevance when we discuss the imaginary contributions to the Euler-Heisenberg Lagrangian.

The Euler-Heisenberg Lagrangian (EHL)\,\cite{Heisenberg:1936nmg,Weisskopf:1936hya} is the most commonly used framework to study the dynamics of constant external background fields. The EHL describes non-linear corrections to the classical Maxwellian Lagrangian in the case of constant electromagnetic backgrounds due to matter field vacuum fluctuations.  The EHL can be organized into several ``loop corrections” to the classical electromagnetic action, each of which is an order $\alpha\equiv q^2/4\pi$ suppressed relative to the previous term (see, for instance,\,\cite{DUNNE_2005,HUET_2012} for a review). The complete EHL may be written formally as
\begin{equation}
    \mathcal{L}_{\text{HE}} = - \frac{1}{4}F^{\mu \nu} F_{\mu \nu} + \sum_{l=1}^{\infty} \mathcal{L}^{(l)} \; .
\end{equation}
The EHL, its extensions, applications and related ideas in broader quantum field theoretic contexts have all been fertile areas of focused research\,\cite{DUNNE_2005,HUET_2012,Dunne:2012vv,Gies:2016yaa,Dumlu_2010}.

In the presence of constant magnetic field backgrounds, say, note that each loop correction $\mathcal{L}^{(l)}$ is a function of two independent quantities--- $\alpha$ and $qB/m^2$, where $B=\lvert\vec{B}\rvert$ is the magnitude of the applied constant magnetic field. Each of the corrections appearing in the EHL may, in the case of weak fields, be expanded as a power series in $q B/m^2$
\begin{equation}\label{eq:general-weak-field-expansion}
    \mathcal{L}^{(l)} \propto m^4 \alpha^{l-1} \sum_{n=0}^{\infty} a_n^{(l)} \left( \frac{qB}{m^2}\right)^{2n} \ , \ \frac{qB}{m^2} \to 0 \; .
\end{equation}

The weak-field expansion diverges for all values of $qB/m^2$, both in the cases of (spinor) QED and sQED (see, for instance,\,\cite{Huet:2017ydx}). The lack of convergence of the weak-field expansion signals the presence of tunnelling instabilities in the theory--for instance, Schwinger pair-production in the large electric field regimes. Generally, such tunnelling instabilities are exponentially suppressed contributions that cannot be directly captured by perturbative expansions of the kind in eq.\,(\ref{eq:general-weak-field-expansion}) and require resurgent extrapolations into trans-series of the form eq.\,(\ref{eq:transs}).

Ideas from resurgence and Borel resummation have recently been brought to bear on analysing spinor QED\,\cite{florio, dunne_higher-loop_2021,non-linear-trident-resummation,Torgrimsson:2022ndq,Dunne:2022esi,Torgrimsson:2021wcj,Torgrimsson:2021zob}. In the present work we hope to extend the resurgent extrapolation studies to the scalar QED EHL, which has not been studied in detail so far in the literature. There have only been a few preliminary comments in the literature pertaining to these topics\,\cite{Huet:2017ydx}. Since sQED is a relatively well-understood quantum field theory, it is an ideal framework to investigate the characteristics of resurgent extrapolation from asymptotic data in the form of weak-field expansions. Analogues of sQED, for instance, involving a dark-sector $U(1)$ gauge field, may also be potential prototypes for models in the dark matter sector\,\cite{HOLDOM1986196,doi:10.1146/annurev.nucl.012809.104433}. These make the investigations here very germane. 

We explore various theoretical aspects and subtleties of the Pad\'{e}-Borel resummation as applied to sQED EHL weak-field expansions at 1-loop and 2-loop order.  We will also contrast the differences and similarities in the sQED EHL analysis with the QED case\,\cite{florio, dunne_higher-loop_2021}, especially in terms of the resurgent structures. In our analysis of the 2-loop sQED EHL, we will also include the recently discovered contribution of one-particle-reducible (1-PR) diagrams\,\cite{Gies:2016yaa}. These have been relatively lesser studied in recent works on the subject. We derive a closed form for these and compare them to the Pad\'{e}-Borel reconstructed functions. Additionally, we will also re-calculate, in modern notation, the formal expressions for the 1-loop and 2-loop particle-irreducible (1-PI) diagrams well-known in the literature\,\cite{v_i_ritus_connection_1977}, and study the resurgent structures appearing in their weak-field expansions. It was recently shown that the EHL for QED loses its meromorphic properties and has a non-trivial trans-series structure at the order of 2-loops\,\cite{dunne_higher-loop_2021}. Additionally, it was shown that methods of resurgence such as Pad\'{e}-Borel and Pad\'{e}-Conformal-Borel transformations can be used to deal with these non-meromorphic properties in the EHL of QED\,\cite{dunne_higher-loop_2021}. We will use the Pad\'{e}-Borel resummation method to demonstrate a similar loss of meromorphic properties in both the 1-PI and 1-PR contributions to the 2-loop EHL of sQED. Additionally, we will demonstrate that the loss of meromorphic properties in the 1-PR contributions has surprising implications for the non-perturbative behaviour of the 2-loop EHL. We will also show that while there are simple relations between the 1-loop EHL of QED and sQED, and the 1-loop sQED EHL and 2-loop 1-PR sQED EHL, the resurgent structures do not seem to follow these simple relations and have non-trivial differences in the singularity structures.

Our work is organized as follows-- In Sec.\,\ref{sec:ehl}, we briefly review the basic ideas and definitions related to EHL, while also fixing notations and conventions. We follow this by deriving the formal expressions for the 1-loop and 2-loop EHL for sQED. We give an overview of the derivation of 2-loop 1-PI contribution to the sQED EHL in Appendix.\,\ref{app_sec:1PI-effective-action}. Then, starting from the weak-field expansions, we use the methods of Pad\'{e}-Borel resummation to reconstruct and theoretically investigate the 1-loop and 2-loop 1-PI and 2-loop 1-PR functions in Sec.\,\ref{sec:pade-borel-analysis-of-ehl}. Specifically, we compare the reconstructed expressions to closed analytic forms in the case of 1-loop and 2-loop 1-PR contributions. We compare the resurgent extrapolations to results from numerical integration for the 2-loop 1-PI contributions since the formal expressions are significantly more intricate and cannot be computed exactly. We derive the closed form for the 1-loop and 2-loop 1-PR EHL for sQED in Appendix.\,\ref{app:closed-form-of-integral}. In particular, we will also investigate the trans-series structure at the one and two-loop orders and study the implications of the loss of meromorphic properties in the 2-loop EHL for sQED. We summarise the main results and conclude in Sec.\,\ref{sec:conclusions}.

\section{Euler-Heisenberg Lagrangian of scalar quantum electrodynamics}\label{sec:ehl}

In this section, we will introduce and review some of the pertinent concepts related to the study. We will also explicitly state all the conventions and definitions used to derive the 1-loop and 2-loop EHLs and related expressions to study resurgence in sQED. While the 1-loop EHL for sQED is relatively well understood from a modern field theoretic perspective, the derivation of the 2-loop contributions remains relatively obscure with a few subtleties. Therefore, we will rederive the expressions while pointing out some of the subtleties, thereby consolidating the pertinent expressions in our adopted conventions. In addition, we will derive novel representations for the two-loop 1-PR EHL for sQED, which have escaped attention until very recently, both as a Schwinger proper time functional integral as well as in a closed analytic form.

In the work, henceforth, $\mathcal{L}_{\text{sQED}}$ will refer to the sQED EHL, and $\mathcal{L}_{\text{QED}}$ will refer to the QED EHL (i.e., with spinors). Whenever the subscript is sometimes omitted in an expression, for brevity, it is then always to be assumed that we are talking about sQED, the main case of interest to us in the paper. We adopt a convention where the flat spacetime metric $\eta_{\alpha\beta}\equiv(1,-1,-1,-1)$. 

The EHL describes the dynamics of a constant electromagnetic background field generated by an external source. The action corresponding to the EHL is most commonly expressed as a classical part with ``loop corrections" due to vacuum fluctuations of the scalar field progressively added to it. This may be formally expressed as
\begin{equation}
    \Gamma_{\text{HE}}[A] = \int d^4x \ \left(- \frac{1}{4} F_{\mu \nu}F^{\mu \nu} \right) + W[A] \; ,
\end{equation}
where $F_{\mu \nu} = \partial_\mu A_\nu - \partial_\nu A_\mu$. The loop action $W[A]$ is the vacuum-persistence amplitude which, in the path integral formalism, can be expressed as
\begin{multline}\label{eq:EHAction_def}
    \exp(i W[A]) = \\ \int \mathcal{D}\widetilde{A} \ \exp \left[i \int d^4x \ \left( - \frac{1}{4} \widetilde{F}_{\mu \nu}^2 + \right. \right. \\ \left. \left. \mathcal{L}_\phi[A+\widetilde{A}] \right) \right] \; .
\end{multline}
where the integration variable $\widetilde{A}$ denotes fluctuations of the photon field and $\widetilde{F}^{\mu \nu}=\partial^\mu \widetilde{A}^\nu-\partial^\nu\widetilde{A}^\mu $. Also, $\mathcal{L}_\phi$ is the Wilsonian effective Lagrangian, with the scalar field integrated out, defined as
    \begin{multline}\label{eq:Wilsonian_Lagrangian}
    \exp \left(i \int d^4x \ \mathcal{L}_\phi[A] \right) = \\ \int \mathcal{D}\phi \mathcal{D}\phi^* \ \exp \left[i \int d^4 x    \ \bigg( (D_\mu \phi)^* (D^\mu \phi) - \right.  \\  \left. m^2 \phi^* \phi \bigg)\right] \; .
\end{multline}
Here, $D_\mu = \partial_\mu+iqA_\mu$ is the covariant derivative. 

It can be shown that the EHL action in eq.\,(\ref{eq:EHAction_def}) may be rewritten as\,\cite{dittrich_effective_1985}
\begin{multline}\label{eq:EHAction_perturbative_form}
        \exp(i W[A]) = \\ \exp \left( \frac{1}{2} \frac{\delta}{\delta J} D_+ \frac{\delta}{\delta J} \right) \exp \left(i \overline{W}^{(1)}[A+J] \right) \Bigg|_{J=0} \; ,
\end{multline}
$\overline{W}^{(1)}$ being the Wilsonian action defined as
\begin{equation}
    \overline{W}^{(1)}[A] = \int d^4x \ \mathcal{L}_\phi[A] \; ,
\end{equation}
$D_+^{\mu \nu} = g^{\mu \nu} D_+(x-y)$ is the photon propagator, and the operator shorthand is defined as
\begin{equation}
     \frac{\delta}{\delta J} D_+ \frac{\delta}{\delta J} = \int d^4x \, d^4y \ \frac{\delta}{\delta J(y)} D_+(x-y) \frac{\delta}{\delta J(x)} \; .
\end{equation}

Expanding the right-hand side in eq.\,(\ref{eq:EHAction_perturbative_form}) gives successive loop corrections to the Euler-Heisenberg action. Let us comment briefly on a few aspects. The functional dependence of $\overline{W}^{(1)}$ on $A$ is expressed through the combination $qA$. As a consequence, every functional derivative of $\overline{W}^{(1)}[A+J]$ with respect to $J$ gives a multiplicative factor of $q$. Consequently, each successive loop correction is an order $q^2$ more than the previous one. Furthermore, as $W[A]$ is represented as the logarithm of the quantity in the right-hand side of eq.\,(\ref{eq:EHAction_perturbative_form}), all the loop corrections only have connected diagrams. Nevertheless, interestingly, this does not rule out the existence of connected 1-PR diagrams at higher orders. While particle-reducible diagrams are not present at the 1-loop order, it has recently been demonstrated\,\cite{Gies:2016yaa} that the contribution of 1-PR diagrams at the two-loop order is non-zero and must be accounted for properly to obtain the complete answer. The work further intimates that the 1-PR contribution, in fact, overtakes the contribution from the 1-PI counterpart in the strong-field limit and, hence, cannot be overlooked in this limit of pertinent interest. With this motivation, we derive explicit representations for the two-loop 1-PR EHL in sQED for the first time.

In Secs.\,\ref{ssec:1-loop-EHL} and \,\ref{ssec:two-loop-EHL}, we will re-express the expressions for the one and two-loop contributions to the EHL in terms of the dressed scalar propagator $G_A(x-y)$ in the background electromagnetic field. The dressed propagator in the proper time representation\,\cite{Schwinger_1951} may be utilised to find the integral representations for the one and two-loop EHLs. 

In Sec.\,\ref{sec:pade-borel-analysis-of-ehl}, for studying resurgent structures in sQED, we will then arrive at functional approximations to these EHL corrections at the one and two-loop orders. These will be obtained leveraging Pad\'{e}-Borel resummations of their weak-field expansions having just a finite number of terms. In this fashion, we may contrast the resummed function constructed from finite-term weak-field expansions to the full integral expression. Specifically, for instance, the closed forms for the 1-loop and 1-PR EHLs may be analytically calculated, making it possible to cross-check the validity of the Pad\'{e}-Borel functional approximations in both the weak and the strong field limit analytically. While a closed analytic form for the 1-PI effective action is not yet known, the function's behaviour in the strong field limit has been analytically calculated in the literature\,\cite{v_i_ritus_connection_1977}. In this case, we will use numerical computations along with the strong field limit expression to check the robustness of the Pad\'{e}-Borel reconstructions. 

\subsection{1-Loop sQED EHL} \label{ssec:1-loop-EHL}

\begin{figure*}[b]
\centering
\includegraphics[width=0.7\linewidth]{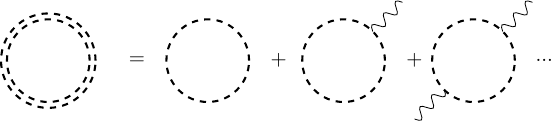}
\caption{A depiction of the 1-loop EHL in the Schwinger proper time framework in terms of the dressed scalar propagator. In terms of explicit Feynman diagrams, it may be considered to include all contributions with one closed scalar loop and any number of background photon insertions.}
\label{sfig:1-loop-feynman-full}
\end{figure*}

\begin{figure*}[b]
\begin{subfigure}{0.27\linewidth}
\centering
\includegraphics[width=0.5\linewidth]{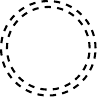}
\caption{}
\label{sfig:1-loop-feynman}
\end{subfigure}
\begin{subfigure}{0.27\linewidth}
\centering
\includegraphics[width=0.5\linewidth]{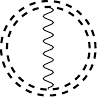}
\caption{}
\label{sfig:two-loop-1pi-feynman}
\end{subfigure}
\begin{subfigure}[b]{0.27\linewidth}
\centering
\includegraphics[width=1.35\linewidth]{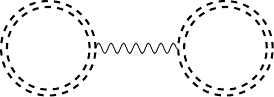}
\caption{}
\label{sfig:two-loop-1pr-feynman}
\end{subfigure}
\caption{Diagrammatic representations of the three contributions to the 1-loop and 2-loop EHL in sQED. Fig.\,\ref{sfig:1-loop-feynman} is the only contribution to the 1-loop sQED EHL, whereas figs.\,\ref{sfig:two-loop-1pi-feynman} and \ref{sfig:two-loop-1pr-feynman} are respectively the 1-PI and 1-PR contributions to the 2-loop sQED EHL. The double dashed lines, as before, denote dressing from the background photon interactions.}
\end{figure*}

Let us start by broadly reviewing the well-known 1-loop contribution to the sQED EHL from a perspective that will be amenable to our resurgence analyses later. The 1-loop correction corresponds to the first term in the expansion of eq.\,(\ref{eq:EHAction_perturbative_form}). Analyzing eq.\,(\ref{eq:EHAction_perturbative_form}), it becomes evident that the 1-loop Euler-Heisenberg action is represented just by $\overline{W}^{(1)}$; as already augured by the notation adopted. Hence, at the level of 1-loop, the EHL is equivalent to the Wilsonian effective action. This, for instance, facilitates the use of the 1-loop EHL to determine light-by-light scattering amplitudes\,\cite{Martin:2003gb,Edwards:2018vjd}. In practice, strictly speaking, a closed analytic form for the 1-loop EHL has only been obtained for constant electromagnetic backgrounds. This makes the utilisation of this particular result for the calculation of scattering amplitudes viable only in the low-frequency limit\,\cite{Dicus_1998}. The 1-loop contribution is conventionally represented by a single loop of dressed propagator, as shown in figs.\,\ref{sfig:1-loop-feynman-full} and\,\ref{sfig:1-loop-feynman}.

One may attempt to find an expression formally for $\mathcal{L}_{\phi}$ in the background electromagnetic field by taking a functional derivative of eq.\,(\ref{eq:Wilsonian_Lagrangian}). This gives us
\begin{multline}\label{eq:1-loop-noether-current}
    \pdv{\mathcal{L}_\phi}{A_\mu} \ (x) =  \\ -iq\bra{A} \phi^*(x) (D^\mu \phi(x)) - \phi(x) (D^\mu \phi(x))^* \ket{A} \; .
\end{multline}
The right-hand side above is just the Noether current corresponding to the $U(1)$ symmetry evaluated in the background electromagnetic field. Particularly, note that the expectation value of the Noether current is being taken, therefore, not with respect to the vacuum $\ket{\Omega}$, but with respect to a state with a background electromagnetic field $A$ present. We have demarcated this state by $\ket{A}$.

The Noether current in sQED is notably distinct compared to its counterpart in spinor QED. This, interestingly, makes for a more intricate calculation of the expectation value in eq.\,(\ref{eq:1-loop-noether-current}), compared to spinor QED. This intricacy is especially pronounced when carrying out the calculation within the framework of the Schwinger proper time formalism (see, for instance,\,\cite{schwartz2014quantum} for a pedagogical discussion) which often leads to this calculation being carried out via functional determinants. However, we will explicitly carry out this calculation in the Schwinger proper time formalism. 

We begin the calculation by expressing eq.\,(\ref{eq:1-loop-noether-current}) in terms of scalar field Green's functions. By using translational invariance in eq.\,(\ref{eq:1-loop-noether-current}), or equivalently from partial integration of the action in eq.\,(\ref{eq:Wilsonian_Lagrangian}), we arrive at the form
\begin{equation}\label{eq:1-loop-noether-current-limit-form}
     \pdv{\mathcal{L}_\phi}{A_\mu } \ (x)  = -2iq \lim_{x \to y} D^\mu_x G_A(x-y) \; ,
\end{equation}
where $G_A(x-y)$ is the scalar field Green's function in the background electromagnetic field. This Green's function, in the Schwinger proper time formalism, may be expressed as 
\begin{multline} \label{eq:greens-function-proper-time}
    G_A(x-y) = \bra{A}\widehat{T} \left[\phi(x) \phi^*(y)\right] \ket{A}  \\ = \int_0^\infty ds \ e^{-s \epsilon} e^{-ism^2} \bra{x} e^{-i\widehat{H}s} \ket{y} \; .
\end{multline}
$\widehat{T}$ denotes time ordering, and $\widehat{H}=-(\widehat{p}-q\widehat{A})^2$, where $\widehat{p}^\mu$ is the momentum operator, and $\widehat{A}^\mu = A^\mu(\widehat{x})$. 

Operating both sides of eq.\,(\ref{eq:greens-function-proper-time}) by $D^\mu_x$ gives rise to a total partial derivative on the right-hand side such that
\begin{multline}
    D^\mu_x G_A(x-y) = \\ \frac{1}{2q} \frac{\partial}{\partial A_\mu}\int_0^\infty \frac{ds}{s} \ e^{-s \epsilon} e^{-ism^2} \bra{x} e^{-i\widehat{H}s} \ket{x} \; .
\end{multline}
Substituting this result in eq.\,(\ref{eq:1-loop-noether-current-limit-form}) gives the well-known 1-loop result for sQED in the Schwinger proper time formalism
\begin{equation}
    \mathcal{L_\phi}(x) = -i \ \int_0^\infty \frac{ds}{s} \ e^{-s \epsilon} e^{-ism^2} \bra{x} e^{-i\widehat{H}s} \ket{x} \; .
\end{equation}
This is illustrated in fig.\,\ref{sfig:1-loop-feynman-full}. Furry's theorem or charge conjugation invariance in sQED will render contributions with an odd number of external photon legs vanishing. Similar derivations of the EHL in the proper time formalism for Lorentz-violating theories have been explored recently\,\cite{Mariz:2007gf,Borges:2016uwl,Ferrari:2021eam}.

As we shall see, the above integral expression for the 1-loop sQED EHL has a much more tractable analytic form as compared to the corresponding 2-loop contributions. In fact, one can even derive a closed analytic form for the 1-loop sQED contribution in terms of the Hurwitz-Zeta functions $\zeta_H(\nu,\alpha)$\,\cite{titchmarsh1986theory,apostol1998introduction}. This gives (see Appendix \ref{app:closed-form-of-integral} for details)
\begin{multline}\label{eq:1-loop-closed-form}
    \mathcal{L}^{(1)}(qB,m^2) = - \frac{m^4}{16 \pi^2} \left(\frac{2qB}{m^2} \right)^2 \times \\ \left[\zeta'_H\left(-1,\frac{1}{2} + \frac{m^2}{2qB}\right) + \frac{3}{4}\left(\frac{m^2}{2qB} \right)^2 \right. \\ \left. +  \left(1 + \log \left(\frac{m^2}{2qB} \right) \right) \zeta_H \left(-1, \frac{1}{2} + \frac{m^2}{2qB} \right) \right] \; .
\end{multline}
The Hurwitz-zeta functions $\zeta_H(\nu,\alpha)$ above are defined as
\begin{equation}
\zeta_H(\nu,\alpha)=\sum_{n=0}^{\infty}\,\frac{1}{(n+\alpha)^\nu}\; ,
\end{equation}
with $\Re(s)>1$ and $\alpha \in +\mathbb{Z}$. $\zeta'_H(\nu,\alpha)$ denotes differentiation with respect to $\nu$.
\subsection{2-Loop sQED EHL} \label{ssec:two-loop-EHL}

Let us now discuss the corresponding 2-loop contributions broadly. The 2-loop sQED EHL can be found by expanding the exponentials in eq.\,(\ref{eq:EHAction_perturbative_form}) up to the second order. This gives
\begin{multline}
    1 + i W^{(1)}[A] + i W^{(2)}[A]+\ldots = \\ \left[1 + \frac{1}{2} \frac{\delta}{\delta J}D_+ \frac{\delta}{\delta J}  \right] \left[1 + i \overline{W}^{(1)}[A] + i J \frac{\delta \overline{W}^{(1)}[A]}{\delta A}  \right. \\ \left. + \frac{iJ^2}{2} \frac{\delta^2 \overline{W}^{(1)}[A]}{\delta A^2} - \frac{J^2}{2} \left(\frac{\delta \overline{W}^{(1)}[A]}{\delta A} \right)^2\right]\Bigg|_{J=0}  \\ + \mathcal{O}(\alpha^2) \; ,
\end{multline}
where we have used the suggestive shorthand 
\begin{equation}
    J \frac{\delta \overline{W}^{(1)}[A]}{\delta A} = \int d^4x \ J(x) \frac{\delta \overline{W}^{(1)}[A]}{\delta A(x)} \; ,
\end{equation}
and so on. 

Only two terms at the second order survive in this expansion. They are illustrated in figs.\,\ref{sfig:two-loop-1pi-feynman} and\,\ref{sfig:two-loop-1pr-feynman}. One of them corresponds to a one-particle-irreducible (1-PI) contribution, and the other to a one-particle-reducible (1-PR) contribution. Specifically, these contributions may be expressed as
\begin{equation}\label{eq:two-loop-1-PI:intial_expression}
    W^{(2)}_{\text{1-PI}}[A] = \frac{1}{2} \int d^4x d^4y \ \frac{\delta^2 \overline{W}^{(1)}}{\delta A^\mu(y) \delta A^\nu (x)} D^{\mu \nu}_+(x-y) \; ,
\end{equation}
and 
\begin{equation}\label{eq:two-loop-1-PR:initial_expression}
    W^{(2)}_{\text{1-PR}}[A] = \frac{i}{2} \int d^4x d^4y \ \frac{\delta \overline{W}^{(1)}}{\delta A^\mu(x)} \frac{\delta \overline{W}^{(1)}}{\delta A^\nu(y)} D^{\mu \nu}_+(x-y) \; .
\end{equation}

The 1-PI effective action contribution at 2-loop order for QED and sQED was first calculated by Ritus\,\cite{v_i_ritus_connection_1977} and has since been recalculated in the worldline formalism\,\cite{Schubert_2001}. To clarify discussions pertinent to the study of resurgence in sQED, we rederive, in our conventions and notations, the relevant expression in the Schwinger proper time formalism (Please see Appendix\,\ref{app_sec:1PI-effective-action})). The 2-loop 1-PI sQED contribution is given formally by the integral
\begin{multline}
    W^{(2)}_{\text{1-PI}}[A] = -iq^2 \int d^4x d^4y  \bigg[4i \delta^4(x-y) \bra{x} \widehat{G}_A \ket{y} \\ - \bra{x} \widehat{\Pi}^\mu \widehat{G}_A \ket{y} \bra{y} \widehat{\Pi}_\mu \widehat{G}_A \ket{x} - \\ \bra{x} \widehat{\Pi}^\mu \widehat{G}_A \widehat{\Pi}_\mu \ket{y} \bra{y}\widehat{G}_A \ket{x}  \bigg] D_+(x-y) \; ,
\end{multline}
where the operators $\widehat{G}_A$ and $\widehat{\Pi}^{\mu}$ are defined as
\begin{subequations}
    \begin{align}
        &\bra{x}\widehat{G}_A\ket{y} = G_A(x-y) \;, \\
        &\widehat{\Pi}^\mu = \widehat{p}^\mu - q \widehat{A}^\mu \; .
    \end{align}
\end{subequations}

While the 1-PI effective action has been extensively studied in various formalisms, the study of the 1-PR contribution has been largely overlooked. It was previously assumed erroneously that this contribution to the QED and sQED EHL was vanishing. It was only recently\,\cite{Gies:2016yaa} that it was demonstrated that the 1-PR contribution in QED is, in fact, non-zero. This realization led to subsequent studies being done in calculating the integral forms for the 1-PR contributions to the propagators and 2-loop EHLs both for QED and sQED\,\cite{Ahmadiniaz_2019,EDWARDS2017339}. In fact, it has been shown\,\cite{Evans_2023} that the 1-PR contribution to the EHL is necessary to account for the everlasting polarization caused by the constant electromagnetic field in space. The closed form for the 1-PR contributions to the 2-loop EHL in QED have been previously calculated\,\cite{Gies:2016yaa}, but a similar calculation has been lacking for the case of sQED.

For sQED, the 2-loop 1-PR contribution may be formally expressed as\footnote{We thank Stefan Evans for pointing out an additional factor of $1/2$ \,\cite{Evans_2023} which was previously missing in the expression for the 1-PR contribution\,\cite{Gies:2016yaa}.}
\begin{equation}\label{eq:two-1-PR-gies_karbstein_expression}
    W^{(2)}_{\text{1-PR}}[A] = \frac{1}{2} \int d^4x \ \pdv{\mathcal{L}^{(1)}}{F^{\mu \nu}} \pdv{\mathcal{L}^{(1)}}{F_{\mu \nu}} \; .
\end{equation}
Although it may seem that the reconstruction of $\mathcal{L}^{(2)}_{\text{1-PR}}$ should readily follow from the reconstruction of $\mathcal{L}^{(1)}$, it is not readily apparent to us that derivatives and multiplications will not affect the resurgent structure of the reconstructed form. In fact, we will see that while $\mathcal{L}^{(1)}$ only has simple poles, the reconstruction of $\mathcal{L}^{(2)}_{\text{1-PR}}$ has a much richer resurgent structure.

The formal expressions for the 1-PI and 1-PR diagrams can be simplified and renormalized to get their expressions as proper time integrals. The proper time integrals for 1-PI diagrams in sQED can be found in a representation similar to their counterparts in QED\,\cite{v_i_ritus_connection_1977}. The 1-PI diagrams in the EHL of QED for a pure magnetic field background have been written as a double integral\,\cite{Lebedev-Ritus-Virial-rep,dunne_higher-loop_2021}. We found a similar representation for sQED in the case of a pure magnetic field background, and it is given by the intricate double integral (see Appendix \ref{app_sec:1PI-effective-action} for a sketch of the derivation)--
\begin{equation}\label{eq:two-loop-1-PI-integral-form}
    \mathcal{L}^{(2)}_{\text{1-PI}} =  \frac{q^2 m^4}{256 \pi^4} \left( \frac{qB}{m^2} \right)^2 \int_0^\infty \frac{dt}{t^3} \ e^{-tm^2/qB}  (T_1 + T_2 + T_3) \; .
\end{equation}
Here, $T_1$, $T_2$ and $T_3$ are given by
\begin{multline}
    T_1 = -\frac{m^2t}{qB} \int_0^1 \frac{du}{u(1-u)} \left[\frac{1}{a_s-b_s} \log \left( \frac{a_s}{b_s} \right) - \right. \\ \left. \frac{t^2}{6} u(1-u) - \frac{t}{\sinh t}  \right] \; , 
\end{multline}
\begin{multline}
    T_2 =  \int_0^1 \frac{du}{u(1-u)} \left[\frac{a_s-c_s}{(a_s-b_s)^2}\log \left(\frac{a_s}{b_s}\right) \right. \\ \left. - \frac{b_s-c_s}{b_s(a_s-b_s)}    +\frac{11t^2}{6}u(1-u)  - \frac{t(1+t\coth t)}{2 \sinh t}  \right] \; ,
\end{multline}
\begin{multline}
    T_3 = \left[1 - \frac{3tm^2}{qB} \left(\log \left(\frac{\gamma tm^2}{qB}\right) - \frac{7}{6} \right)\right] \times \\ \left[\frac{t}{\sinh t} - 1 + \frac{t^2}{6}  \right] \; ,
\end{multline}
with
\begin{equation}
    a_s =  \frac{\sinh(tu)\sinh(t(1-u))}{t^2u(1-u)} \; ,
\end{equation}
\begin{equation}
    b_s = \frac{\sinh t}{t} \; ,
\end{equation}
\begin{multline}
    c_s = \cosh(tu) \cosh(t(1-u)) - \\ 3 \sinh(tu) \sinh(t(1-u)) \; .
\end{multline}
$\log\gamma$ is the Euler's constant; $\log \gamma \sim 0.577$\,\cite{abramowitz1965handbook}. 
Here, we have added the subscript `s' to the variables $a_s$, $b_s$, and $c_s$ to differentiate them from the values of $a$, $b$, $c$ typically defined in the 1-PI EHL for QED. Note that while the overall structure of the 1-PI integrals for QED and sQED is the same, the functional form of $T_1$, $T_2$, and $T_3$ are very different from their counterparts in QED. This is very much unlike the case of the 1-loop EHL, where we can relate the QED and sQED EHLs with a simple equation as in eq.\,(\ref{eq:1-loop-scalar-spinor-relation}). 

The rather complicated double integral structure of the two-loop EHL given for QED in\,\cite{Lebedev-Ritus-Virial-rep} and the one we derived in eq.\,(\ref{eq:two-loop-1-PI-integral-form}) leads to the presence of branch cuts in the two-loop EHL, in contrast to the one-loop EHL given in eq.\,(\ref{eq:1-loop:integral-form}). As a consequence, the imaginary contribution of the two-loop EHL has a much more intricate structure than that of the one-loop EHL. This intricate structure manifests as the branch-cut function $K_n(\beta,s)$ in the imaginary contributions to the EHL's of QED and sQED, where $\beta = qE/m^2$ and $s=0,1$ for sQED and QED respectively. Although the analytic analysis of the function $K_n(\beta,s)$ has been done in detail\,\cite{Lebedev-Ritus-Virial-rep}, this analysis is quite cumbersome and certainly not practical for higher loop orders. The recent effort of resurgence has proven to be useful in predicting the properties of such branch cuts without having to perform any complicated analytic analysis. Such a resurgent analysis has been performed for two-loop QED\,\cite{dunne_higher-loop_2021} and a similar analysis for sQED is presented in Sec.\,\ref{ssec:two-loop-PB}.

The rather complicated double integral of $\mathcal{L}^{(2)}_{\text{1-PI}}$ has no closed form. The 1-PR contribution, on the other hand, is relatively straightforward. It may be evaluated starting from eq.\,(\ref{eq:two-1-PR-gies_karbstein_expression}), adapting the methodologies detailed in\,\cite{Gies:2016yaa}. Proceeding along these lines, for the 2-loop sQED EHL 1-PR contribution, one gets the double-integral expression (for a constant magnetic field background)--
\begin{equation}
    \mathcal{L}^{(2)}_{\text{1-PR}} = \frac{q^4 B^2}{1024 \pi^4} I_2^2 \; .
\end{equation}
Here, $I_2$ denotes the derivative of the 1-loop integral and is given by
\begin{equation}\label{eq:two-loop-1PR-integral}
    I_2 = \int_0^{\infty} \frac{ds}{s^3} \ e^{-m^2s/qB} \left[\frac{s}{\sinh s}-\frac{s^2}{\sinh s \tanh s} + \frac{s^2}{3} \right] \; .
\end{equation}
In fact, we were able to derive a closed form for $I_2$ using methods detailed in Appendix \ref{app:closed-form-of-integral}--
    \begin{multline}
    I_2 = -4 \zeta_H \left(-1,\frac{1+m^2/qB}{2} \right) - \\ 8 \zeta_H' \left(-1,\frac{1+m^2/qB}{2} \right) + \\  2 \left(\frac{m^2}{qB}\right) \zeta_H' \left(0,\frac{1+m^2/qB}{2} \right)  -\frac{1}{3} \log \left(\frac{m^2}{2qB} \right) \; .
\end{multline}
In the following section, we will now use these integral forms for the 1-loop, 2-loop 1-PI, and 2-loop 1-PR contributions to the sQED EHLs to investigate the resurgent characteristics in sQED. For instance, we would like to understand how well one may reconstruct the respective sQED EHL contributions, given their weak-field asymptotic expansions up to some finite number of terms.

\section{Pad\'{e}-Borel analysis of the sQED EHL} \label{sec:pade-borel-analysis-of-ehl}

In many physical situations, one is faced with perturbative series that are occasionally factorially divergent and, therefore, don't converge anywhere on the complex plane. Dyson\,\cite{dyson} first noted the significance of such divergences and argued heuristically that a divergent series may be an indication of non-perturbative behaviour in some sectors of the theory---for instance, signifying non-perturbative instabilities manifested by instantons. That the nature of the non-perturbative sectors may already be gleaned from the perturbative sectors is among the lessons from the broad program of resurgence (see, for example,\,\cite{Aniceto:2018bis,Costin:2020hwg}, and references therein). Particularly, in many cases, the methods of Pad\'{e}-Borel analysis (see, for example,\,\cite{Costin:2019xql,pade-borel-review,marino_2015}) may help one re-construct the almost complete functional form from just a few terms of the small-coupling or weak-field perturbative expansions. 

In this section, we explore resurgent extrapolation in sQED as pertaining to its large-field and weak-field behavior. The weak-field expansions of the sQED EHL are factorially divergent at both one and two-loop orders, as will be shown in Secs.\,\ref{ssec:1-loop-PB} and \ref{ssec:two-loop-PB}. We investigate in detail how efficiently we can reconstruct large-field characteristics of the theory from just a finite number of terms of the weak-field expansions. In this context, let us start by discussing some of the concepts in resurgent extrapolation we require for our analyses. 

Consider a factorially divergent asymptotic series $\mathcal{S}$ with the coefficients $a_n$,
\begin{equation}\label{eq:def-formal-divergent-series}
    \mathcal{S}(z) = \sum_{n=0}^{\infty} a_n z^{2n} \; ,
\end{equation}
where the factorial divergence is of the form
\begin{multline}\label{eq:general-large-order-behaviour}
    a_n \sim \sum_{\mathcal{P}=1}^{\infty}C_\mathcal{P} \ A_\mathcal{P}^{-2n-k_{\mathcal{P}}-1} \ \Gamma(2n+k_\mathcal{P}+1) + \\ \text{corr.} \ , \ n \to \infty \;.
\end{multline}
Here, $k_\mathcal{P}$ is a fixed constant. Every successive term in the sum of eq.\,(\ref{eq:general-large-order-behaviour}) is exponentially smaller than the next, since $A_1< A_2 < A_3 \dots$ and we refer to them as ``power-law contributions" (see, for instance,\,\cite{dunne_higher-loop_2021,marino_2015}). The corrections in eq.\,(\ref{eq:general-large-order-behaviour}) refer to sub-leading corrections to each power-law contribution. As was mentioned previously, depending on the quantity under consideration, factorial divergences may be manifestations of instabilities in some sectors of the theory---they may be related to sub-dominant saddle points of the action of the theory. As we shall see, our weak-field expansions will have features similar to the asymptotic series $ \mathcal{S}(z) $.

When the function underlying $\mathcal{S}$ is reconstructed using methods of Borel summation, these instabilities are manifested as exponentially suppressed imaginary contributions to the function and are therefore missed by a naive analysis of the perturbation series in eq.(\ref{eq:def-formal-divergent-series}). The idea of resurgence is to relate these exponentially suppressed imaginary contributions to the large order behaviour of $a_n$-- in fact, in the particular case, it dictates that each power-law contribution gives an exponentially suppressed contribution proportional to $e^{-|A_P/z|}$. The study of these and other features in trans-series is an active area of mathematical research\,\cite{trans-series-research1,trans-series-research2,trans-series-research-3,Schiappa:2023ned}. In the physics context, exponentially suppressed terms of this form are associated with instanton contributions--- for example, the leading order contribution to the imaginary part $e^{-|A_1/z|}$ is associated with the first instanton contribution.

In the context of QED and sQED, the values of $\{ A_\mathcal{P} \}$ are often imaginary, indicating instanton contributions to the imaginary part of $\mathcal{S}$ for imaginary values of $z$. The Borel dispersion relations, which have been expounded upon in Appendix\,\ref{app:borel-dispersion-relations}, may then be used to quantify the above notion and explicitly compute the instanton contributions. These instanton contributions combine to form a trans-series for the imaginary part of $\mathcal{S}$ such that, (see, for instance,\,\cite{trans-series-ref1,trans-series-ref2} and references therein),
\begin{equation}\label{eq:trans-series-def}
    \text{Im} \ \mathcal{S}(it) = \sum_{\mathcal{P}=1}^{\infty} \frac{ i \pi C_{\mathcal{P}}}{2 (it)^{k_{\mathcal{P}}+1}} \ e^{-A_{\mathcal{P}}/t} + \text{corr.} \; ,
\end{equation}
The index $\mathcal{P}$ in the sum represents the $\mathcal{P}^{\text{th}}$ instanton contribution. Corrections to the power law causes each instanton contribution to be an independent asymptotic series; this is a characteristic of a trans-series. 

In the following sub-sections, we will denote the $\mathcal{P}^{\text{th}}$ instanton contribution by $\mathcal{I}_{\mathcal{P}}$. Therefore, in this notation, we can also rewrite eq.\,(\ref{eq:trans-series-def}) as
\begin{equation}\label{eq:def-instantons}
    \text{Im} \ \mathcal{S}(it) = \sum_{\mathcal{P}=1}^{\infty} \mathcal{I}_{\mathcal{P}}^{(\mathcal{S})}(t) \; .
\end{equation}
The process of Pad\'{e}-Borel reconstruction begins by defining the Borel transform. We will define a general Borel transform $\widehat{\mathcal{S}}_{p,N^*} (z)$ of the asymptotic series $ \mathcal{S}(z)$ as
\begin{equation}
    \widehat{\mathcal{S}}_{p,N^*} (z) = \sum_{n=0}^{N^*} \frac{a_n}{(2n+p)!} \ z^{2n+p} \ , \ p\geq 0 \; .
    \label{eq:BTdef}
\end{equation}
In the limit $N^*\to \infty$, the series $\widehat{\mathcal{S}}$ converges within the radius of convergence $A_1$. 

For finite $N^*$, the next step in the analysis is to find an analytical continuation for the Borel transform such that it reproduces the properties of $\widehat{\mathcal{S}}$ inside and on the radius of convergence while also furnishing a satisfactory analytical continuation outside the radius of convergence. The analytical continuation generally used for the single-variable Borel transforms are the Pad\'{e} approximants (see, for instance,\,\cite{baker1975essentials,baker1996pade}). These are rational functions denoted by $P^{N}_{M}(x)$, with the parameters $N$ and $M$ representing the degrees of the polynomials in the numerator and the denominator.  

In our specific case, the Pad\'{e} approximants will be constructed to have the same weak-field expansion inside the radius of convergence as the weak-field Borel transform. Specifically, in terms of $\widehat{\mathcal{S}}$, we equate
\begin{multline}\label{eq:def-pade-approximant}
    P^N_{M}\big[\widehat{\mathcal{S}}\big](z)\equiv \frac{U_0 + U_1 z + \dots + U_Nz^N}{1 + D_1 z + \dots D_M z^M} \\ \sim \sum_{n=0}^{N^*} \frac{a_n}{(2n+p)!} z^{2n+p}\equiv  \widehat{\mathcal{S}}_{p,N^*} (z) \, ,~\ z \to 0 \; ,
\end{multline}
where $N^*$ denotes the number of coefficients required to construct $P^N_M$ ($N^*$ will depend on $N$, $M$ and $p$). The diagonal and off-diagonal Pad\'{e} approximants, corresponding to $M=N$ and $M=N+1$, are the most frequently used approximants in literature\,\cite{dunne_higher-loop_2021,florio}. We will denote the Pad\'{e} approximants in use by $P^{N}_{N+f}$, where the parameter $f$ is a measure of how far the Pad\'{e} approximant is from the diagonal Pad\'{e} approximant. 

The Borel sum of the formal series $\mathcal{S}$ is defined as a Laplace like transform of $\widehat{S}$ such that (see, for instance,\,\cite{guillou1990large} and references therein)
\begin{equation}\label{eq:def-borel-transform}
    \text{B}_p\big[\mathcal{S} \big](z) = z^{-p}\lim_{N^* \to \infty} \int_0^{\infty} dt \ e^{-t} \ \widehat{\mathcal{S}}_{p,N^*}(zt) \; .
\end{equation}
It follows from a simple use of the definition of the Gamma function, $\Gamma(n+1) = \int_0^{\infty} dt \ e^{-t} \ t^{n}$\,\cite{abramowitz1965handbook}, that we get 
\begin{equation}
    \mathcal{S}(z) \sim \text{B}_p\big[\mathcal{S} \big](z) \ , \ z \to 0 \; .
\end{equation}
We can make the method of Borel sum better by constructing Pad\'{e} approximants of the Borel transform $\widehat{\mathcal{S}}$ out of only finite terms $N^*$ and taking a Laplace-like transform to get the ``reconstructed function" $\text{PB}_{N,f,p}\big[\mathcal{S} \big]$. The Pad\'{e}-Borel reconstruction is defined as
\begin{equation}\label{eq:def-PB-recontruction}
    \text{PB}_{N,f,p}\big[ \mathcal{S} \big](z) = z^{-p} \int_0^{\infty} dt \ e^{-t} \ P^N_{N+f}\big[\widehat{\mathcal{S}}_{p,N^*}\big](zt) \; .
\end{equation}
The parameters $f$ and $p$ are often omitted in conventional definitions of Pad\'{e}-Borel reconstructions. Nonetheless, it is important to note that varying these parameters can change the properties of the reconstructed function in meaningful ways. The apriori unknown optimal values of $f$ and $p$ can be explored by looking at the asymptotic behaviour of the function to be reconstructed in the limit of strong fields, as we will see in Secs.\,\ref{ssec:1-loop-PB} and \ref{ssec:two-loop-PB}. 

The perturbative information obtained from the asymptotic series in eq.\,(\ref{eq:def-formal-divergent-series}) is insufficient to uniquely reconstruct the underlying function $\mathcal{S}$ without additional information. Given just a perturbative series as in eq.\,(\ref{eq:general-weak-field-expansion}), it is possible to construct a range of functions that are equally effective for small values of $z$ but exhibit different behaviour for larger values of $z$. This relates to the concept of ``non-perturbative ambiguity", which is typically quantified by the non-perturbative error $e^{-|A_1/z|}$ (see, for instance,\,\cite{marino_2015}). This error represents the extent to which $\mathcal{S}$ can be approximated. In Secs.\,\ref{ssec:1-loop-PB} and\,\ref{ssec:two-loop-PB}, we will use different schemes of $f$ and $p$ and explicitly demonstrate the non-perturbative ambiguity that accompanies the reconstruction of asymptotic series using the methods of Pad\'{e}-Borel reconstruction.

In some cases like QED and sQED, obtaining some information or data points regarding the behaviour of the function underlying $\mathcal{S}$ might be possible for large values of $z$ by doing some non-perturbative analysis. A unique functional approximation for $\mathcal{S}$ can be fixed in such cases. 

In the following section, we will consider a finite number of terms in the weak-field expansion of the sQED EHL at 1-loop and 2-loop orders and carefully analyze the resurgent extrapolation in each case. Additionally, we will show how a unique functional approximation can be fixed by using the large-field behaviour of the 1-loop and 2-loop EHLs, which can be derived by comparing the trace anomalies of the EHL with the full theory at $\mathcal{O}(\alpha^2)$.

\subsection{At the 1-loop order}\label{ssec:1-loop-PB}

Let us first focus on the 1-loop sQED EHL expression and the ensuing weak-field expansion in the presence of a magnetic field. While resurgent properties of the 1-loop EHL for QED, which is given by\,\cite{Heisenberg:1936nmg}
\begin{multline}
    \mathcal{L}^{(1)}_{\text{QED}}(qB,m^2) = \\  -\frac{q^2B^2}{8 \pi^2} \int_0^{\infty} \frac{dt}{t^2} \ e^{-m^2t/qB} \left[\coth t  - \frac{1}{t} - \frac{t}{3} \right] \; ,
\end{multline}
is relatively well studied\,\cite{DUNNE_2005,florio}, similar studies for the sQED EHL remain unexplored. We will start the Pad\'{e}-Borel analysis by looking at the integral form of the sQED EHL at 1-loop order in the presence of a constant magnetic field background, given by (see, for instance,\,\cite{pady,DUNNE_2005})
\begin{multline}\label{eq:1-loop:integral-form}
    \mathcal{L}^{(1)}_{\text{sQED}} (qB,m^2) = \\ \frac{q^2 B^2}{16 \pi^2} \int_0^{\infty} \frac{dt}{t^2} \ e^{-m^2t/qB} \left[ \frac{1}{\sinh t} - \frac{1}{t} + \frac{t}{6} \right] \; .
\end{multline}
We note, by comparing to the spinor QED case above, that eq.\,(\ref{eq:1-loop:integral-form}) can be represented in terms of the 1-loop EHL of spinor QED, using the identity $1/\sinh t = \coth \left(t/2 \right) - \coth (t)$. Specifically, the correspondence one obtains is\,\cite{Weisskopf:1936hya}
\begin{multline}\label{eq:1-loop-scalar-spinor-relation}
    \mathcal{L}^{(1)}_{\text{sQED}} (qB,m^2) = \\ \frac{1}{2}\mathcal{L}^{(1)}_{\text{QED}} (qB,m^2) - \frac{1}{4}\mathcal{L}^{(1)}_{\text{QED}} (qB,2m^2)  \; .
\end{multline}
Although it may seem that this simple relation can be used to get a reconstruction of the sQED EHL using corresponding reconstructions of the QED EHL, we will see in our analysis that there are some differences in the resurgent structure of the 1-loop QED and sQED EHLs. This seems to indicate a loss of additivity in resurgent structures in this context. 

One may obtain a weak field expansion from the sQED 1-loop EHL in eq.\,(\ref{eq:1-loop:integral-form}). This is of the form
\begin{multline}
    \mathcal{L}^{(1)}_{\text{sQED}}(qB,m^2) \sim  \frac{m^4}{4 \pi^2} \left(\frac{qB}{m^2} \right)^4 \sum_{n=0}^{\infty} a_n^{(1)} \left(\frac{qB}{m^2} \right)^{2n} \\ \equiv  \frac{m^4}{4 \pi^2} \left(\frac{qB}{m^2} \right)^4 \mathcal{S}^{(1)}_{\text{sQED}}(qB/m^2) \ , \ ~ \frac{qB}{m^2} \to 0 \; ,
\end{multline}
where the coefficients $a_n^{(1)}$ may be found exactly, and have the form
\begin{equation}\label{eq:1-loop-coefficient}
    a_n^{(1)} = (-1)^n \ \frac{2^{2n+3}-1}{(2\pi)^{2n+4}} \ \zeta (2n+4) \ (2n+1)! \; ,
\end{equation}
and the zeta function is defined as $\zeta(\nu) = \sum_{n=1}^{\infty}1/n^\nu$. The relative simplicity of sQED, along with the factorial divergence of the coefficients of the weak-field expansion of the sQED EHL, makes it among the ideal theoretical cases to explore the efficacy of Pad\'{e}-Borel reconstructions. We will omit the subscripts from $\mathcal{L}_{\text{sQED}}$ and $\mathcal{S}_{\text{sQED}}$, as we will only talk about the EHL and weak-field expansions for sQED.

We define the general Borel transform of $\mathcal{S}^{(1)}$ appearing in the 1-loop sQED EHL as 
\begin{equation}
    \widehat{\mathcal{S}}_{p,N^*}^{(1)} (g_B) = \sum_{n=0}^{N^*} \frac{a_n^{(1)}}{(2n+p)!} \ g_B^{2n+p} \ , \ p\geq 0 \; ,
\end{equation}
where we have defined $g_B=qB/m^2$, a dimensionless measure of the strength of the external magnetic field. Henceforth, we will occasionally use this notation for brevity of expressions wherever required. 

The reconstruction is done by approximating the Borel series $\widehat{\mathcal{S}}_{p,N^*}^{(1)}$ by Pad\'{e} approximants and then performing a Laplace-like transform, as described in eq.\,(\ref{eq:def-PB-recontruction}). This furnishes a functional approximation for $\mathcal{L}^{(1)}$.
\begin{multline}\label{eq:temp0911231}
    \text{PB}_{N,f,p}\big[ \mathcal{S}^{(1)}\big] (g_B) = \\ g_B^{-p}\int_0^{\infty} dt \ e^{-t} \ P^{N}_{N+f}\big[ \widehat{\mathcal{S}}^{(1)}_{p,N^*}\big](g_B t) \; .
\end{multline}

To investigate the effectiveness of the Pad\'{e}-Borel resummation, we compare the reconstructed function $\text{PB}_{p,N,f} \mathcal{S}^{(1)}$ to the closed form, as given in eq.\,(\ref{eq:1-loop-closed-form}), of $\mathcal{L}^{(1)}$ for arbitrary field values
\begin{multline}\label{eq:temp2610231}
    \mathcal{L}^{(1)}(qB,m^2) \xLeftrightarrow{~~?~~} \\ \frac{m^4}{4 \pi^2} \left(\frac{qB}{m^2} \right)^{4} \text{PB}_{p,N,f} \big[ \mathcal{S}^{(1)}\big] (g_B) \; .
 \end{multline}
\begin{figure*}[t!]
    \centering
    \includegraphics[width=0.9\linewidth]{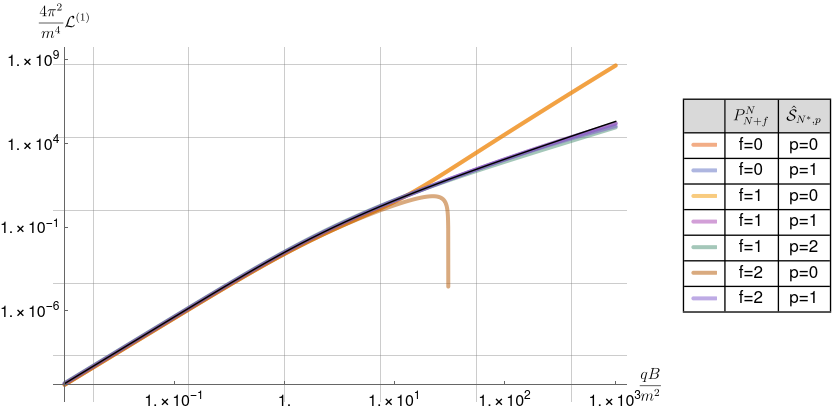}
    \caption{A comparison of different parameter values $p$ and $f$ and their effectiveness at replicating the original function. It is apparent that the parameter combinations $p=1$, $f=1$ provide the best functional approximation.}
    \label{fig:L1-diff-mp}
\end{figure*}

To get a sense of what the optimal values for the apriori unknown parameters $f$, $p$ may be, let us consider the leading large-field behaviour of $\mathcal{L}^{(1)}$. This may be found by comparing the trace anomalies of the EHL\,\cite{DUNNE_2005} with the full theory, at first order in $\alpha$. For sQED, this gives
\begin{equation}
    \mathcal{L}_{\text{sQED}}^{(1)} \sim \frac{q^2B^2}{96 \pi^2} \log \left(\frac{qB}{m^2} \right) \ , ~\frac{qB}{m^2} \rightarrow \infty \; .
\end{equation}
In passing, we note that the corresponding expression for QED which although it has the same logarithmic growth, has a different pre-factor due to differences in the value of the coefficients of their respective beta functions\,\cite{DUNNE_2005}. It is given by
\begin{equation}
    \mathcal{L}^{(1)}_{\text{QED}} \sim \frac{q^2B^2}{24 \pi^2} \log \left(\frac{qB}{m^2} \right) \ , \ \frac{qB}{m^2} \to \infty \; .
\end{equation}
We may compare the large-field behaviour of $\mathcal{L}_{\text{sQED}}$ with that of the Pad\'{e}-Borel reconstructed Lagrangian, which is given by 
\begin{multline}\label{eq:1-loop-PB-large-order-behaviour}
  \frac{m^4}{4\pi^2} \left(\frac{qB}{m^2} \right)^{4} \text{PB}_{N,f,p}\big[ \mathcal{S}^{(1)}\big] (g_B) \sim \\ \frac{m^4}{4\pi^2} \left(\frac{qB}{m^2} \right)^{4-p} \int_{1/g_B}^{\infty} dt \ e^{-t} \ \frac{U_N/D_{N+f}}{(g_Bt)^f}  \ , ~\ g_B \to \infty \; ,
\end{multline}
where $U_N$ and $D_{N+f}$ are the coefficients of the $N^{\text{th}}$ and $(N+f)^{\text{th}}$ degree terms in the numerator and denominator of the Pad\'{e}-approximant as defined in eq.\,(\ref{eq:def-pade-approximant}). The infrared region of the integral $[0,1/g_B)$ may be ignored in the strong-field limit as it will be $\mathcal{O}(1/g_B)$. The integral in eq.\,(\ref{eq:1-loop-PB-large-order-behaviour}) is the incomplete Gamma function $\Gamma(-f+1,x)$ with $x=1/g_B$, which is defined as\,\cite{abramowitz1965handbook}
\begin{equation}
    \Gamma(a,x) = \int_x^{\infty} dt \ e^{-t} \ t^{a-1}\;.
\end{equation}
The derivative of the incomplete Gamma function, $\partial_x \Gamma(-f+1,x) = - \frac{e^{-x}}{x^f}$, can be expanded in the limit $x\to 0$ to deduce that, up to a multiplicative constant, we would have
\begin{multline}
    \frac{m^4}{4 \pi^2} \left(\frac{qB}{m^2} \right)^{4} \text{PB}_{N,f,p} \big[ \mathcal{S}^{(1)}\big] \sim \\ \frac{m^4}{4 \pi^2} \left(\frac{qB}{m^2} \right)^{4-p} \begin{cases}
\frac{g_B^{f-1}}{g_B^f}, & \text{if } f > 1 \\
\frac{\log g_B}{g_B} & \text{if } f=1 
    \end{cases} ~\ ,~~~ \frac{qB}{m^2} \to \infty \; .
\end{multline}
From the above discussions, we deduce that the optimal parameter choices would be $p=1$ and $f=1$.
\begin{figure*}[t!]
    \centering
    \includegraphics[width=0.9\linewidth]{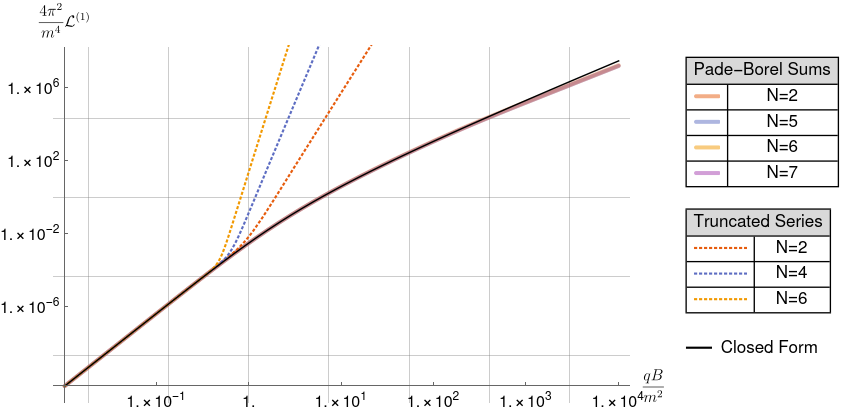}
    \caption{A comparison of the effectiveness of the weak-field asymptotic expansion and the Pad\'{e}-Borel reconstruction at approximating $\mathcal{L}^{(1)}$. The dashed lines represent the asymptotic expansion truncated at different values of $N$. The coloured lines represent the Pad\'{e}-Borel reconstructions with Pad\'{e} approximants of the form $P^N_{N+1}$ for different values of $N$, and the black line represents the closed form. The generalized Borel sum is used here with the parameter values $p=1$, $f=1$.}
    \label{fig:L1-Asy-vs-Pade}
\end{figure*}

We see a similar trend in fig.\,\ref{fig:L1-diff-mp}-- the superior approximation is given by $p=1$, $f=1$. We were able to predict that these parameter values would give the optimum approximation because we had some information about the behaviour of $\mathcal{L}^{(1)}$ in the specific case of sQED for large $g_B$ due to non-perturbative arguments (a similar thing also holds for QED; please refer to\,\cite{dunne_higher-loop_2021}). However, we would not know how to choose the optimal parameter values without this information. Without non-perturbative information, we obtain a set of functional approximations corresponding to different parameter values $p$ and $f$, as in fig.\,(\ref{fig:L1-diff-mp}). While all of these approximations work equally well in the perturbative sector, they differ for values of $g_B \sim 10$, indicative of a non-perturbative ambiguity. 

In fig.\,\ref{fig:L1-Asy-vs-Pade}, we compare the Pad\'{e}-Borel reconstruction of the 1-loop sQED finite-termed weak-field expansion with the full 1-loop sQED EHL. We have taken $f=1$ and $p=1$ here. Even with just two terms ($N=2$) from the truncated weak field expansion, we see that the Pad\'{e}-Borel reconstruction is still able to well-approximate the full 1-loop sQED EHL in the strong-field regime; up to $g_B \sim 10^3$.

Now, the above analysis was performed for a constant magnetic field background. The analytical continuation $B \to \pm iE$ can extend this analysis to constant electric field backgrounds. The motivation for such a rotation stems from the Lorentz invariant quantity $B^2-E^2$, on which the EHL should always depend. Our analysis will always choose the analytical continuation $B\to -iE$. We provide a heuristic argument for our reason for doing so; other arguments have also been presented in the literature\,\cite{jentschura-gies}. 

\begin{figure}[t!]
\centering
\includegraphics[width=\linewidth]{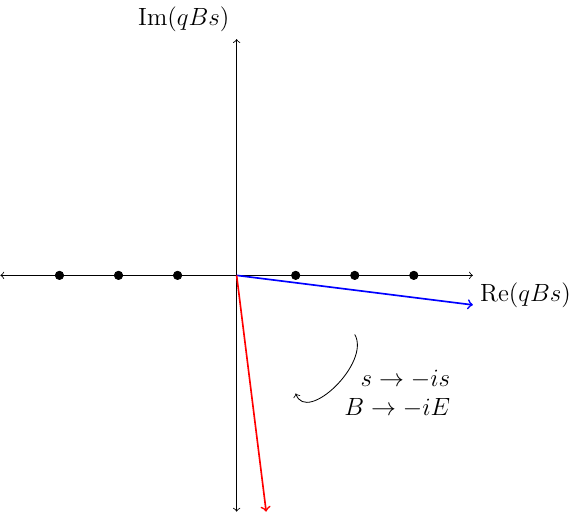}
\caption{Valid analytical continuations in the $qBs$ plane. The continuations $s\to -is$ and $B \to -iE$ are equivalent.}
\label{fig:analytical-continuation}
\end{figure}

Consider, for example, the 1-loop EHL for sQED. Its most preliminary functional form is given by\,\cite{pady}
\begin{equation}
    \mathcal{L}^{(1)}_{\text{sQED}} = - \int_0^{\infty} \frac{ds}{s^3} e^{-i(m^2-i\epsilon)s} F(qBs) \; ,
\end{equation}
where the function $F$ is given by
\begin{equation}
    F(x) = \frac{x}{\sin x} - 1 - \frac{x^2}{6} \; .
\end{equation}
The $\epsilon$ prescription is necessary to ensure convergence of the integral. This prescription causes the integration contour to lie slightly below the real axis, as shown in fig.\,\ref{fig:analytical-continuation}. We note here that because the functional dependence of $s$ always comes in the form of $qBs$, this coordinate is a more natural choice to work in. To compute this integral in the presence of magnetic fields, the integration variable $s$ is traditionally continued to the negative imaginary axis, i.e., $s\to -is$. Here, the continuation from $s$ to $is$ is not viable due to the existence of poles along the real axis. But because the functional dependence of $s$ comes in the form of $qBs$, the analytic continuation $B \to iE$ should not be viable either.

\begin{figure*}[t!]
    \begin{subfigure}{0.55\linewidth}
        \includegraphics[width=\linewidth]{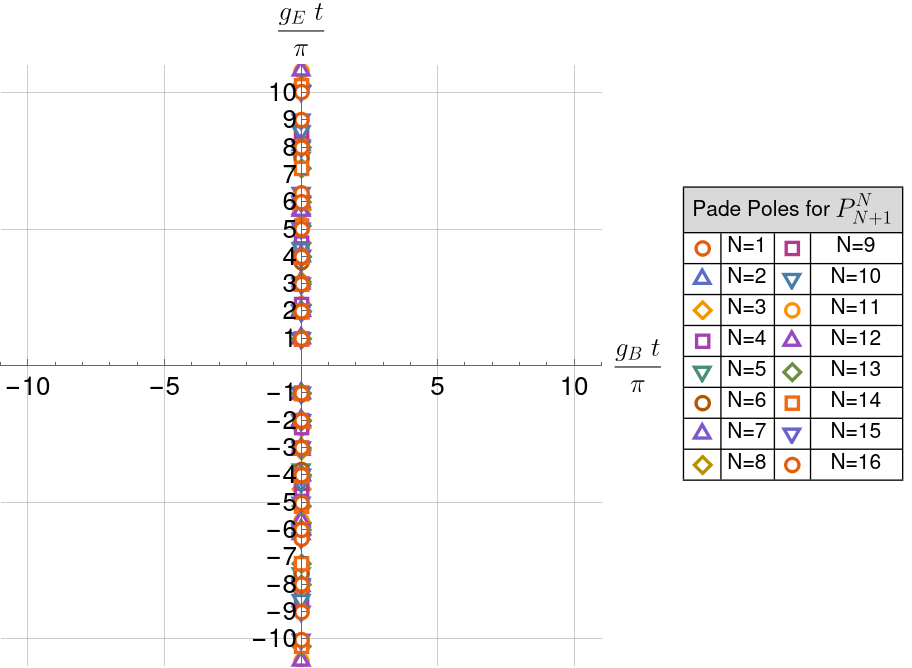}
        \caption{}
        \label{sfig:L1-pole-normal}
    \end{subfigure}
    \begin{subfigure}{0.35\linewidth}
        \includegraphics[width=1.1\linewidth]{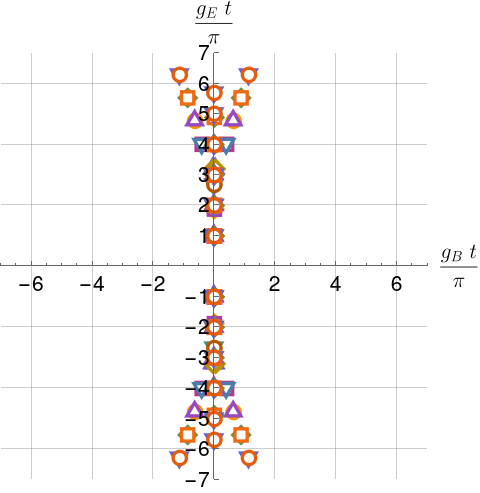}
        \caption{}
        \label{sfig:L1-pole-szego}
    \end{subfigure}
    \caption{Poles of the Pad\'{e} approximants of the 1-loop EHL constructed in two different ways. A naive construction of Pad\'{e} approximants leads to arcs of spurious poles as seen in fig.\,\ref{sfig:L1-pole-szego}, whereas Pad\'{e}-like approximants constructed from the spinor 1-loop EHL do not give rise to such spurious poles, as is seen in fig. \ref{sfig:L1-pole-normal}. }
    \label{fig:L1-poles}
\end{figure*}

However, when the analytic continuation $g_B \to -ig_E$ is implemented on the Pad\'{e}-Borel reconstructed function in eq.\,(\ref{eq:def-PB-recontruction}), it is found that the Pad\'{e} approximant $P^N_{N+f}$ develops a series of poles on the imaginary axis, as is shown in fig.\,\ref{fig:L1-poles}. In this case, the singularities of the Pad\'{e} approximant of the Borel transform are simple poles at integral multiples of $i\pi$. This is also expected because of the $1/\sinh$ term in the integrand in the exact expression eq.\,(\ref{eq:1-loop:integral-form}).
In fig.\,\ref{fig:L1-poles}, we have constructed the pole structure of the Pad\'{e}-approximants using two strategies. In the first strategy, we directly construct the Pad\'{e} approximant $P^N_{N+1}$ from $\widehat{\mathcal{S}}^{(1)}$. In the second strategy, we instead compute separately the Pad\'{e}-approximants for the Borel transform series of the 1-loop EHLs of spinor QED, as they appear in eq.\,(\ref{eq:1-loop-scalar-spinor-relation}). From this, we then construct Pad\'{e}-like approximants for the 1-loop EHL of sQED. 

\begin{figure}[t!]
    \centering
    \includegraphics[width=\linewidth]{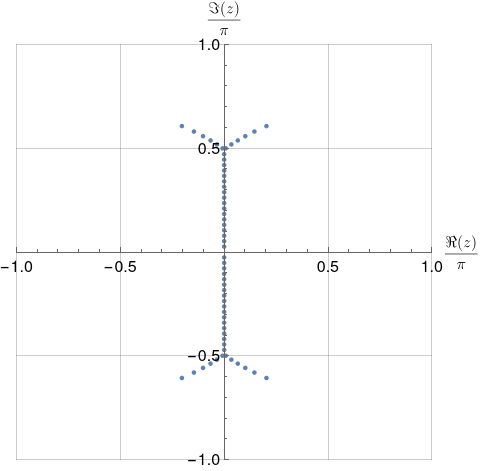}
    \caption{The poles of the normalized Pad\'{e} approximant $P^{60}_{61}(120z)$ illustrating the Szeg\H{o} curves. The plot shows the characteristic curve of the spurious poles of a normalized partial sum $s_n(nz)$, as discussed in the text.}
    \label{fig:L1-szego-curves}
\end{figure}

While both strategies show simple poles on the imaginary axis, we see curiously, by comparing to the full theory, that the first strategy gives rise to unwanted arcs of spurious poles. This may be related to the so-called Szeg\H{o} curves. Loosely speaking, Szeg\H{o} curves are curves onto which the zeros of a normalized partial sum of $e^z$, denoted as $s_n(nz)$ converge (see, for instance, \cite{Szego-1,szego-2}). It has been demonstrated that analogous curves also manifest in the distribution of zeros of the normalized partial sums of $\sin(z)$ and $\cos(z)$ functions\,\cite{varga_zeros_2000}. In our context, for the first strategy, we similarly observe characteristic Szeg\H{o} curves in the distribution of the poles of the normalized Padé approximants. This is depicted in fig.\,\ref{fig:L1-szego-curves}. 

Such spurious poles are interestingly absent in the Pad\'{e}-like approximants constructed starting from the spinor Borel transform series, as is seen in fig.\,\ref{sfig:L1-pole-normal}. Therefore, the second strategy detailed above seems to be able to circumvent the issue with spurious poles. Thus, comparing with the Pad\'{e} approximants derived from the integral in eq.\,(\ref{eq:1-loop:integral-form}), this methodology seems to give a better resurgent extrapolation. Note that the answers obtained from the two strategies should agree in the limit $N \to \infty$, as the Szeg\H{o} curves are artifacts from partial sums. In this limit, the arc of spurious poles is pushed to infinity. This trend is also seen explicitly in fig.\,\ref{sfig:L1-pole-szego}, where the arc of spurious poles is pushed farther and farther away from the origin for larger and larger $N$.

 The presence of poles in the Pad\'{e} approximants seemingly introduces singularities in the integral in eq.\,(\ref{eq:def-PB-recontruction}) because of the presence of poles in the term $P^N_{N+f}(zt)$ for $z=-ig_E$. However, note that if this integral is performed with the analytic continuation $B \to -iE + \epsilon$, where $\epsilon \to 0$, we can circumvent these poles in the integral. This manoeuvre is equivalent to doing the variable change $t \to it/g_E$ (where $g_E = qE/m^2$) and a contour rotation to get that
\begin{multline}\label{eq:def:PB-resum-electric-field}
    \text{PB}_{N,f,p} \big[\mathcal{S}^{(1)} \big] (g_E) = \\ \frac{1}{(-ig_E)^{p+1}} \int_0^{\infty} dt \ e^{-it/g_E} \ P^N_{N+f}\big[\widehat{\mathcal{S}}^{(1)}_{N^*,p} \big] \; .
\end{multline}
This integral will have both a real and an imaginary part. Therefore, an analytical continuation to electric fields gives rise to an imaginary contribution to the EHL. This imaginary contribution is responsible for the false vacuum decay that leads to Schwinger pair production\,\cite{Schwinger_1951}. The instability of the vacuum in the presence of strong electric fields leads to the production of charged particle and anti-particle pairs and hence causes vacuum polarization. This effect is entirely non-perturbative and is suppressed by a factor of $e^{-\pi m^2/qE}$. Therefore, effects such as these cannot be probed by conventional perturbative analysis methods and require methodologies developed in the resurgence program. 

Note that because the poles of $P^N_{N+f}$ only lie on the imaginary axis, we do not encounter singularities while reconstructing the EHL in the integral in eq.\,(\ref{eq:def-PB-recontruction}) in the case of a pure magnetic field background. Therefore, the EHL doesn't develop an imaginary part for a pure magnetic field background.

\begin{figure*}[t!]
    \centering
    \includegraphics[width=0.9\linewidth]{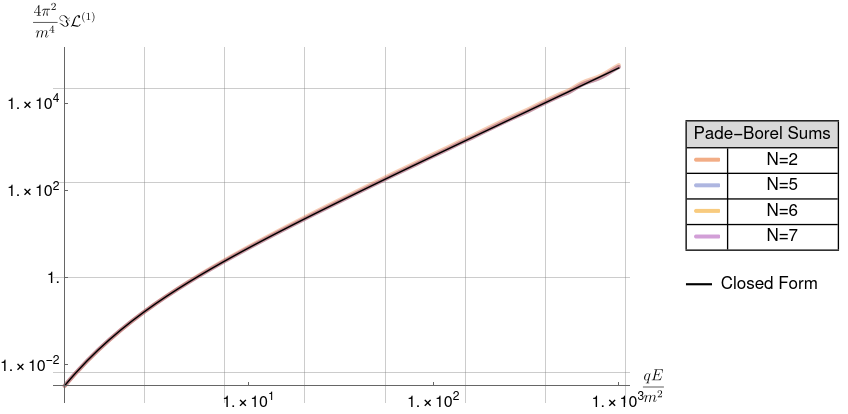}
    \caption{Pad\'{e}-Borel reconstruction of the imaginary of the 1-loop EHL analytically continued to electric fields, compared with the closed form. The parameter values used here are $p=1$, $f=1$}
    \label{fig:L1-reconstruction-electric}
\end{figure*}

We will use the Pad\'{e}-Borel reconstructions to reconstruct the imaginary part of the EHL and show that it gives remarkably close approximations to the full theory, even though the information used to construct it is perturbative.  The results from the scheme of Pad\'{e}-Borel reconstruction as was described in eq.\,(\ref{eq:def:PB-resum-electric-field}) are compared in fig.\,\ref{fig:L1-reconstruction-electric}. It is apparent that the Pad\'{e}-Borel reconstruction method is not only successful in approximating the EHL for magnetic field backgrounds but can also effectively provide very good functional approximations in the electric field regimes. The parameter values used here are again $p=1$, $f=1$.

\begin{figure*}[t!]
    \centering
    \includegraphics[width=0.9\linewidth]{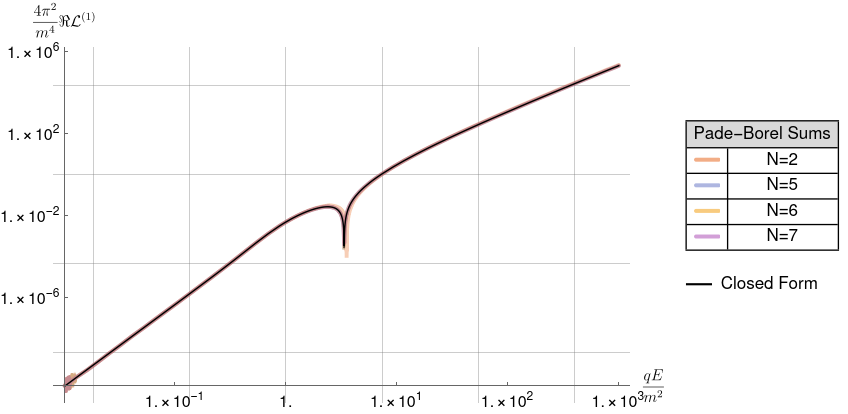}
    \caption{Pad\'{e}-Borel reconstruction of the real part of the 1-loop EHL analytically continued to electric fields, compared with the closed form. The Pad\'{e}-Borel resummations successfully approximate the closed form, even at the kink at $g_E \sim 10$, which marks a transition in the function's sign. The parameter values used here are $p=1$, $f=1$}
    \label{fig:enter-label}
\end{figure*}
As was mentioned at the beginning of Sec.\,\ref{sec:pade-borel-analysis-of-ehl}, the program of resurgence, especially the Borel dispersion relations, can be used to relate the large-order behaviour of weak-field coefficients to the imaginary part of the function underlying the weak-field expansion. If the formal structure of the weak-field expansion is known, we may determine the large order behaviour of the coefficients $a_n^{(1)}$ up to all orders. From eq.\,(\ref{eq:1-loop-coefficient}), we deduce that at large orders
\begin{multline}\label{eq:1-loop-large-order-behaviour-an}
    a_n^{(1)} \sim (-1)^n \ \frac{2^{2n+3}-1}{(2\pi)^{2n+4}} \ \Gamma(2n+2) \left[1 + \frac{1}{2^{2n+4}} + \right. \\ \left. \frac{1}{3^{2n+4}} + \dots \right] \ , \ n \to \infty \; .
\end{multline}
Cognizant of the large order behaviour, we may now compute the exact contribution of each instanton. This can be done by deducing the singularity structure of the Borel transform $\widehat{S}$ using the large-order behaviour of $a_n^{(1)}$ and then using this singularity structure to find its contribution to the imaginary part of the Borel sum $\text{B}[\mathcal{S}]$\,\cite{Dunne:1999vd}. The result of this procedure is known as the Borel dispersion relations, which are detailed in Appendix\,\ref{app:borel-dispersion-relations}.

It is apparent from the structure of the large-n expansion of $a_n$ that the Borel transform $\widehat{\mathcal{S}}^{(1)}$ must have multiple simple pole singularities at integer multiples of $\pi$. This attribute is also realized in the pole structure of the corresponding Pad\'{e} approximants. Already, we see evidence of non-perturbative behaviour lurking in the large order behaviour of $a_n^{(1)}$. From the Borel dispersion relations in Appendix \ref{app:borel-dispersion-relations}, and the large order behavior in eq.\,(\ref{eq:1-loop-large-order-behaviour-an}), we calculate the instanton contributions to the 1-loop EHL to be of the form
\begin{multline}\label{eq:scalar-imaginary-part}
    \text{Im} \ \mathcal{L}^{(1)}_{\text{sQED}} = \frac{m^4}{16 \pi^3} \left( \frac{qE}{m^2} \right)^2 \left[e^{-m^2\pi/qE}  \right. \\ \left. - \frac{1}{4}e^{-2m^2\pi/qE} + \frac{1}{9} e^{-3m^2 \pi/qE} + \dots \right] \; .
\end{multline}
We may compare this with the imaginary part of the spinor QED 1-loop EHL, reconstructed in a similar fashion\,\cite{florio,dunne_higher-loop_2021}
\begin{multline}\label{eq:spinor-imaginary-part}
    \text{Im} \ \mathcal{L}^{(1)}_{\text{QED}} = \frac{m^4}{8 \pi^3} \left( \frac{qE}{m^2} \right)^2 \left[e^{-m^2\pi/qE} + \right. \\ \left. \frac{1}{4}e^{-2m^2\pi/qE} +  \frac{1}{9} e^{-3m^2 \pi/qE} + \dots \right] \; .
\end{multline}
This reproduces the well-known result\,\cite{Lebedev-Ritus-Virial-rep,DUNNE_2005,Kim:2003qp} of the vacuum decay rate for sQED in the regime of strong electric fields. The quantity that is most closely related to observations is the average number of pairs produced per unit volume per unit time. This is just given by the first term in the above series\,\cite{Cohen:2008wz,Ruffini:2009hg}.

\subsection{At the 2-loop order}\label{ssec:two-loop-PB}
Let us now turn our focus on the 2-loop contributions to the sQED EHL and a study of the resurgent extrapolation of the corresponding weak-field expansions. At the 2-loop order, eq.\,(\ref{eq:EHAction_perturbative_form}) suggests, as discussed earlier, that there may be two distinct contributions to the sQED EHL---the one-particle irreducible contribution and a one-particle reducible contribution. It is found that both these contributions are physically significant, with interesting inter-plays between them as the magnitude of the background electromagnetic field increases. Unlike the 1-loop expressions, the 2-loop expressions are relatively less tractable analytically. 

The above formal expression for 1-PI in eq.\,(\ref{eq:two-loop-1-PI-integral-form}) has no known simple closed analytic form, and we will evaluate these numerically for future comparisons. In contrast, the expression for the 1-PR contribution in eq.\,(\ref{eq:two-loop-1PR-integral}) is relatively straightforward. 

We begin the Pad\'{e}-Borel analysis of the 2-loop sQED EHL contributions by performing a weak-field expansion for the 1-PI and 1-PR expressions for a constant magnetic field background. We will denote the coefficients for the 1-PI EHL by $a_n^{(2I)}$, and that of the 1-PR EHL by $a_n^{(2R)}$. The corresponding weak-field expansions have the form
\begin{multline}
    \mathcal{L}_{\text{1-PI}}^{(2)} \sim \frac{q^2 m^4}{256 \pi^4} \left(\frac{qB}{m^2} \right)^4 \sum_{n=0}^\infty a_n^{(2I)} \left(\frac{qB}{m^2} \right)^{2n} \\ \equiv  \frac{q^2 m^4}{256 \pi^4} \left(\frac{qB}{m^2} \right)^4  \mathcal{S}^{(2I)}(qB/m^2)   \ , \ \frac{qB}{m^2} \to  0 \; ,
\end{multline}
\begin{multline}
    \mathcal{L}_{\text{1-PR}}^{(2)} \sim \frac{q^2 m^4}{1024 \pi^4} \left(\frac{qB}{m^2} \right)^6 \sum_{n=0}^\infty a_n^{(2R)} \left(\frac{qB}{m^2} \right)^{2n} \\ \equiv \frac{q^2 m^4}{1024 \pi^4} \left(\frac{qB}{m^2} \right)^6  \mathcal{S}^{(2R)}(qB/m^2)  \ , \ \frac{qB}{m^2} \to 0 \; .
\end{multline}

Due to the more tractable analytic expression as compared to its 1-PI counterpart, it is feasible to numerically compute the exact weak-field coefficients of the 1-PR integral up to the desired orders relatively fast. Conversely, the weak-field expansion of the 1-PI integral is significantly more intricate, leading to quickly escalating computation times. We computed the first 27 coefficients of the weak field expansion of the 1-PI integral and have noted the first 15 of these values in Appendix.\,\ref{app:2-loop-coefficient-values} for reference. The first few of these coefficients in the 2-loop sQED EHL weak-field expansions may be compared to the existing values in literature\,\cite{Ahmadiniaz_2019,kors_effective_1999}, which in our convention may be represented as
\begin{multline}\label{eq:1-PI-weak-field-expansion}
    \mathcal{L}^{(2)}_{\text{1-PI}} \sim \frac{q^2m^4}{256 \pi^4}   \left[\frac{275}{648} \left(\frac{qB}{m^2} \right)^4 -\frac{5159}{16200}\left(\frac{qB}{m^2} \right)^6 + \right. \\ \left. \frac{751673}{1058400} \left(\frac{qB}{m^2} \right)^8 + \dots \right] \ , \ \frac{qB}{m^2} \to 0 \; ,
\end{multline}
\begin{multline}\label{eq:1-PR-weak-field-expansion}
    \mathcal{L}^{(2)}_{\text{1-PR}} \sim \frac{q^2 m^4}{1024 \pi^4}  \left[ \frac{49}{8100} \left(\frac{qB}{m^2} \right)^6 - \frac{31}{2700} \left(\frac{qB}{m^2} \right)^8 + \right. \\ \left. \frac{58433}{1587600} \left(\frac{qB}{m^2} \right)^{10} + \dots \right] \ , \ \frac{qB}{m^2} \to 0 \; .
\end{multline}
The computed values in Appendix.\,\ref{app:2-loop-coefficient-values} are seen to match these values in the existing literature very well.

A ratio test on the weak-field coefficients of the 1-PI integral shows that they scale as
\begin{equation}\label{eq:1-PI-large-order}
    a_n^{(2I)} \sim 8.0 \ (-1)^n \ \frac{\Gamma(2n+2)}{\pi^{2n+2}} \; .
\end{equation}
The weak-field coefficients of the 1-PR integral reveal that, unlike $a_n^{(2I)}$, $a_n^{(2R)}$ scale as
\begin{equation}\label{eq:1-PR-large-order}
    a_n^{(2R)} \sim   0.1 \ (-1)^n \  \frac{\Gamma(2n+3)}{\pi^{2n+3}} \; .
\end{equation}
 The results of the ratio tests on $a_n^{(2I)}$ and $a_n^{(2R)}$ are tabulated in tables.\,\ref{table:a_n^2I-fits} and \ref{table:fit-a_n2R}. Clearly, the growth is factorial, supporting the asymptotic analysis above.

\begin{table}[h]
    \centering
    \begin{tabular}{|c|c|}
    \hline
        $n$ & $an^2+bn+c$ \\
    \hline \hline
       $5-10$  & $0.404191 n^2+0.243097 n+0.128455$\\ \hline
       $10-15$ & $0.404741 n^2+0.232291 n+0.181699$\\ \hline
       $15-20$ & $0.404984 n^2+0.225143 n+0.234264$\\ \hline
       $20-25$ & $0.405094 n^2+0.220784 n+0.277354$ \\
       \hline
    \end{tabular}
    \caption{A tabulation of $a_n^{(2I)}/a_{n-1}^{(2I)}$ fits to a quadratic polynomial $an^2+bn+c$ for different ranges of $n$. The left column depicts the ranges of $n$ for which the ratio $a_n^{(2I)}/a_{n-1}^{(2I)}$ was calculated and fitted, and the right column depicts the fit obtained from these ranges of $n$. The values of the fit-parameters $a$ and $b$ are seen to converge and indicate the asymptotic factorial growth of the 1-PI coefficients $a_n^{(2I)} \sim \Gamma(2n+2)/\pi^{2n}$.}
    \label{table:a_n^2I-fits}
\end{table}
\begin{table}[h]
\centering
    \begin{tabular}{|c|c|}
    \hline
         $n$ & $a n^2+bn+c$ \\
    \hline
    \hline
      $20-30$   & $0.405136 n^2+0.618621 n -0.449305$\\
      \hline
      $30-40$ & $0.405238 n^2+0.612768 n-0.364672 $ \\
      \hline
      $35-45$ & $ 0.405254 n^2 +0.611527 n -0.341554$ \\
      \hline
      $40-50$ &  $0.405264 n^2+0.610714 n -0.324364 $ \\
      \hline
    \end{tabular}
\caption{$a_n^{(2R)}/a_{n-1}^{(2R)}$ ratio fits to the quadratic polynomial $an^2+bn+c$. The values of the fit parameters $a$ and $b$ are again seen to converge, indicating the asymptotic factorial growth of the 1-PR coefficients $a_n^{(2R)} \sim \Gamma(2n+3)/\pi^{2n}$.}
\label{table:fit-a_n2R}
\end{table}

As in the 1-loop case, we perform a Borel transform and construct a Pad\'{e} approximant to the truncated Borel transform series. In this case, we will have two Borel transforms pertaining to the 1-PI and 1-PR diagrams---
\begin{equation}
    \widehat{S}^{(2I)}_{p,N^*}(g_B) = \sum_{n=0}^{N^*} \frac{a_n^{(2I)}}{(2n+p)!} g_B^{2n+p} \; ,
\end{equation}
\begin{equation}
    \widehat{S}^{(2R)}_{p,N^*}(g_B) = \sum_{n=0}^{N^*} \frac{a_n^{(2R)}}{(2n+p)!} g_B^{2n+p} \; .
\end{equation}
The Pad\'{e} approximants are then used in a Laplace-like transform as in eq.\,(\ref{eq:def-PB-recontruction}) to obtain the resurgent extrapolation of the 2-loop 1-PI and 1-PR sQED weak-field expansions. We will then investigate the efficacy of different schemes of Pad\'{e}-Borel resummations in reconstructing the 1-PI and 1-PR contributions to the 2-loop EHL of sQED.
\begin{multline}
    \mathcal{L}^{(2)}_{\text{1-PI}}(qB,m^2) \xLeftrightarrow{~~?~~} \\ \frac{q^2m^4}{256 \pi^4} \left(\frac{qB}{m^2} \right)^{4} \text{PB}_{N,f,p} \big[ \mathcal{S}^{(2I)}\big] (g_B) \; ,
\end{multline}
\begin{multline}
    \mathcal{L}^{(2)}_{\text{1-PR}}(qB,m^2) \xLeftrightarrow{~~?~~}  \\ \frac{q^2 m^4}{1024 \pi^4} \left(\frac{qB}{m^2} \right)^{6} \text{PB}_{N,f,p} \big[ \mathcal{S}^{(2R)}\big] (g_B) \; .
\end{multline}

\begin{figure*}[t]
    \centering
    \includegraphics[width=0.9\linewidth]{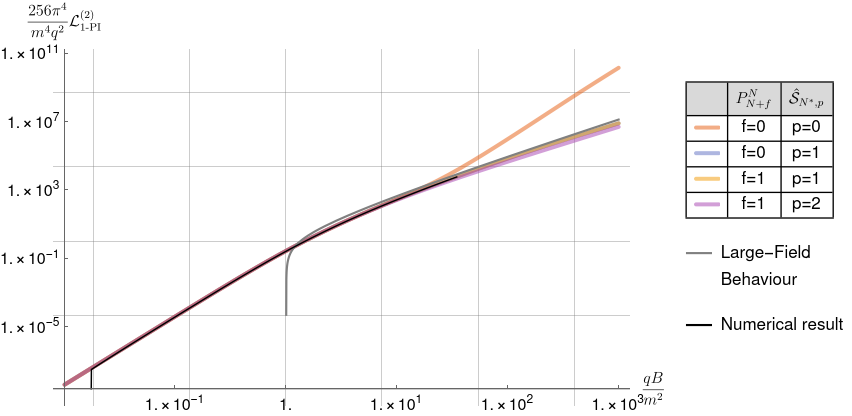}
    \caption{A comparison of the effectiveness of various schemes of Pad\'{e}-Borel approximations, parameterized by different values of $p$ and $f$, in reconstructing $\mathcal{L}^{(2)}_{\text{1-PI}}$ for $N=6$. Clearly, the parameter values $p=1$ and $f=1$ provide the closest approximation to the numerically-computed closed form and the large-order behaviour of $\mathcal{L}^{(2)}_{\text{1-PI}}$.}
    \label{fig:two-loop-1PI-diff-mp}
\end{figure*}

\begin{figure*}[t]
    \centering
    \includegraphics[width=0.9\linewidth]{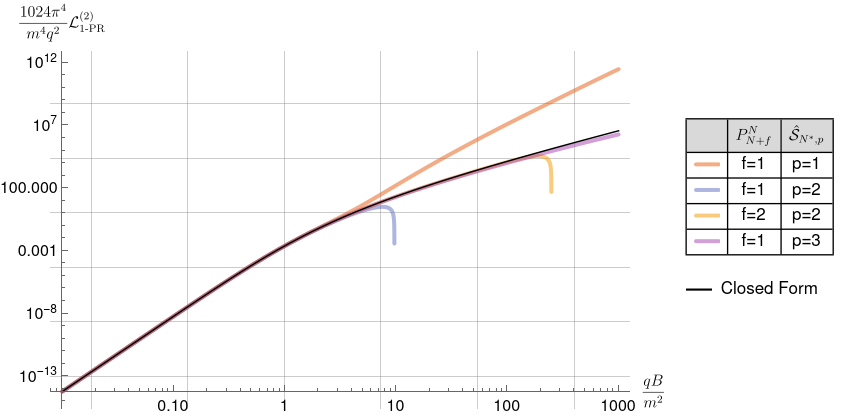}
    \caption{A comparison of the effectiveness of various schemes of Pad\'{e}-Borel approximations, parameterized by different values of $m$ and $p$, in reconstructing $\mathcal{L}^{(2)}_{\text{1-PR}}$ for $N=6$ }
    \label{fig:two-loop-1PR-diff-mp}
\end{figure*}

\begin{figure*}[b]
    \centering
    \includegraphics[width=0.9\linewidth]{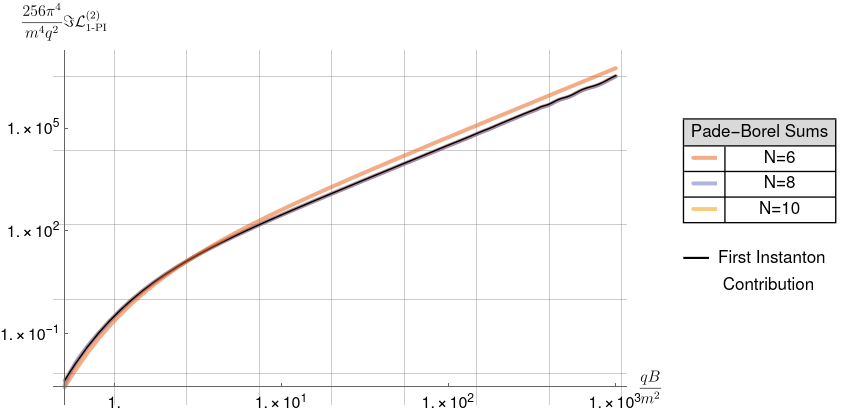}
    \caption{Comparison of the Pad\'{e}-Borel reconstruction of the imaginary part of the 1-PI Lagrangian in the regime of strong electric fields with the first instanton contribution.}
    \label{fig:L2I-electric-field}
\end{figure*}

\begin{figure*}[t]
    \centering
    \includegraphics[width=0.9\linewidth]{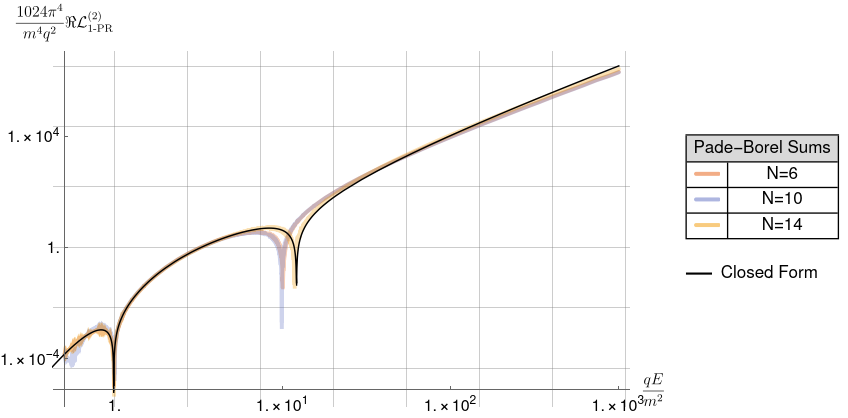}
    \caption{Pad\'{e}-Borel reconstruction of the real part of the analytically continued 1-PR EHL compared with its closed form. The behaviour of the 1-PR EHL in the electric field regime is quite different from that of its 1-loop and two-loop 1-PI counterparts in that it changes its sign twice. The Pad\'{e}-Borel resummation provides an excellent approximation to the closed form, effectively reproducing both of its kinks.  The parameter values used for the reconstruction are $f=1$, and $p=3$}
    \label{fig:L2R-electric-real-part}
\end{figure*}
\begin{figure*}[t]
    \includegraphics[width=0.9\linewidth]{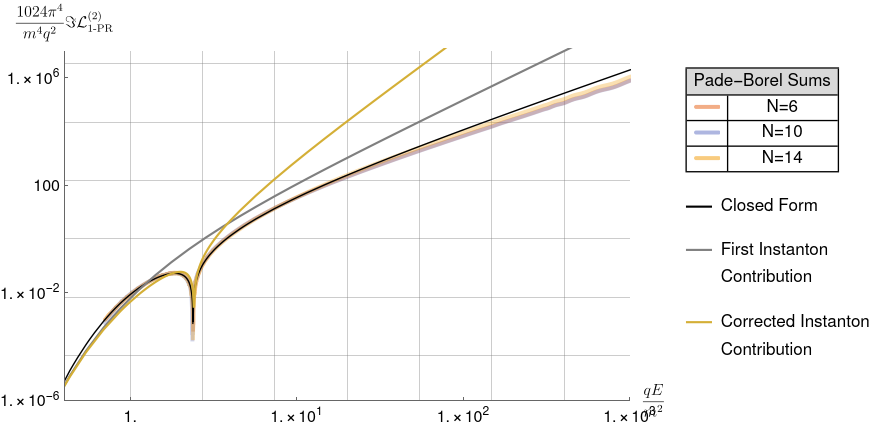}
    \caption{Comparison of the Pad\'{e}-Borel reconstruction of the imaginary part of the analytically continued 1-PR Lagrangian with the closed form. Unlike its 1-loop and two-loop 1-PI counterparts, the imaginary contribution of the 1-PR EHL in the regime of electric fields flips sign, as is signified by the kink in the graph. While the naive estimate of the first instanton contribution (illustrated in grey) is unable to reproduce this kink, a sub-leading correction to the first instanton contribution (illustrated in golden) can explain this behaviour. The parameter values for the Pad\'{e}-Borel approximations are $f=1$, $p=3$}
    \label{fig:L2R-electric-field-imaginary-part}
\end{figure*}
Again, we will use different values of $p$ and different schemes of the Pad\'{e} approximants $P_N^{N+f}$ to construct Pad\'{e}-Borel resummations from the weak field coefficients $a_n^{(2I)}$ and $a_n^{(2I)}$, and further optimize the functional approximation. To get a sense of what the optimal parameter values $p$ and $f$ might be, as defined in eqs.\,(\ref{eq:BTdef}) and (\ref{eq:def-PB-recontruction}), we may again compare the strong-field behaviours, as we did in the previous sub-section. The strong field behaviours for the 1-PI and 1-PR diagrams are known to be\,\cite{DUNNE_2005,karbstein-all-loop}---
\begin{equation}
    \mathcal{L}_{\text{1-PI}}^{(2)} \sim \frac{q^4 B^2}{128 \pi^4} \log \left(\frac{qB}{m^2} \right) \ , ~\frac{qB}{m^2} \rightarrow \infty \; ,
\end{equation}
\begin{equation}
    \mathcal{L}^{(2)}_{\text{1-PR}} \sim  \frac{1}{9} \frac{q^4B^2}{1024 \pi^4} \left[\log \left(\frac{qB}{m^2} \right)\right]^2 \ , ~\frac{qB}{m^2} \rightarrow \infty \; .
\end{equation}

On the other hand, the large-order behaviour of the Pad\'{e}-Borel reconstructed 2-loop Lagrangians will look like 
\begin{multline}
       \frac{q^2 m^4}{256 \pi^4} \left(\frac{qB}{m^2} \right)^4 \text{PB}_{N,f,p} \big[\mathcal{S}^{(2I)}\big]  \sim  \\ \frac{q^2m^4}{256 \pi^4} \left(\frac{qB}{m^2} \right)^{4-p}\begin{cases}
\frac{g_B^{f-1}}{g_B^f}, & \text{if } f > 1 \\
\frac{\log g_B}{g_B} & \text{if } f=1
    \end{cases} \; ,
\end{multline}
and,
\begin{multline}
    \frac{q^2 m^4}{1024 \pi^4} \left(\frac{qB}{m^2} \right)^6 \text{PB}_{N,f,p} \big[\mathcal{S}^{(2R)}\big] \sim \\ \frac{q^2m^4}{1024\pi^4} \left(\frac{qB}{m^2}\right)^{6-p} \begin{cases}
\frac{g_B^{f-1}}{g_B^f}, & \text{if } f > 1 \\
\frac{\log g_B}{g_B} & \text{if } f=1
    \end{cases} \; .
\end{multline}
Comparing the two large-order behaviours, it is apparent that the best approximation would correspond to the parameter values $p=1$, $f=1$ for the 1-PI Lagrangian and $p=3$, $f=1$ for the 1-PR Lagrangian. 

To ascertain that these parameter values indeed give the best functional approximations, we compare the efficacy of various Pad\'{e}-Borel reconstructions---computed using different parameter values of $p$ and $f$. These comparisons are shown in figs.\,\ref{fig:two-loop-1PI-diff-mp} and \ref{fig:two-loop-1PR-diff-mp}. Given that we lack a closed-form expression for the 1-PI EHL, we have compared the Pad\'{e}-Borel reconstructed Lagrangian from $\text{PB}\big[\mathcal{S}^{(2I)}\big]$ with results from the numerical integration of eq.\,(\ref{eq:two-loop-1-PI-integral-form}). A comparison is also shown with the analytic large-field behaviour of $\mathcal{L}^{(2)}_{\text{1-PI}}$. In contrast,  we can find a closed analytic form for the sQED 2-loop 1-PR contribution (see Appendix \ref{app:closed-form-of-integral})--
\begin{multline}\label{two-loop:1-PR-closed-form}
    \mathcal{L}_{\text{1-PR}}^{(2)} = \frac{q^4 B^2}{1024 \pi^4} \left[ -4 \zeta_H \left(-1,\frac{1+m^2/qB}{2} \right) - \right. \\ \left. 8 \zeta'_H \left(-1,\frac{1+m^2/qB}{2} \right) \right. \\ \left. + 2 \left(\frac{m^2}{qB}\right) \zeta'_H \left(0,\frac{1+m^2/qB}{2} \right)  -\frac{1}{3} \log \left(\frac{m^2}{2qB} \right) \right]^2 \; .
\end{multline}
Figs.\,\ref{fig:two-loop-1PI-diff-mp} and \ref{fig:two-loop-1PR-diff-mp} illustrate that optimal and robust functional approximations for the 1-PI and 1-PR contributions are achieved when utilizing parameter values of $p=1$, $f=1$ and $p=3$, $f=1$, respectively. For the optimal parameter values, the 2-loop Pad\'{e}-Borel reconstructed functions are seen to give remarkably accurate functional approximations, even with a truncation at $N=6$.

As was done in the 1-loop case, we can again analytically continue the magnetic field case to the electric field case via the prescription $B \to -iE$. We may then compare the imaginary part of the analytically continued full 1-PI and 1-PR contribution to the instanton contribution computed using eqs.\,(\ref{eq:1-PI-large-order}), (\ref{eq:1-PR-large-order}) and (\ref{app-eq:borel-dispersion-relation}) from the weak-field expansions. 

The Borel dispersion relations, in this case, must be used with care, owing to the different large-order behaviours of the 1-PI and 1-PR weak-field coefficients. The dispersion relations, as is shown in Appendix \ref{app:borel-dispersion-relations}, are a technique to find instanton contributions to the imaginary part of the EHL using the large-order behaviour for weak-field expansions. Given a set of weak field coefficients $a_n$ with the large order behaviour
\begin{multline}\label{eq:temp0811232}
    a_n \sim  \ \sum_{\mathcal{P}=1}^{\infty} C_{\mathcal{P}} \ (iA_{\mathcal{P}})^{-2n-k_{\mathcal{P}}-1} \ \Gamma(2n+k_{\mathcal{P}}+1)  \\ \ , \ n \to \infty \; ,
\end{multline}
the Borel dispersion relations dictate that the imaginary contribution to the EHL be given by
\begin{equation}
    \text{Im}\big[\text{PB} \big[\mathcal{S} \big] \big](g_E) = \frac{i\pi}{2} \sum_{\mathcal{P}=1}^{\infty} \frac{C_{\mathcal{P}}}{(ig_E)^{k_{\mathcal{P}}+1}} e^{-A_{\mathcal{P}}/g_E} \; .
\end{equation}
Making use of these relations for the 1-PI and 1-PR diagrams with $A_1 = \pi$, $C_1 = -8.0$, $k_{1}=1$, and $A_1 = \pi$, $C_1 = -0.1i$, $k_1=2$ respectively, we get the instanton contributions $\mathcal{I}^{(2I)}_1$ and $\mathcal{I}^{(2R)}_1$--
\begin{subequations}
    \begin{align}
        &\mathcal{I}^{(2I)}_1 = \frac{q^2 m^4}{256 \pi^4} \ (-ig_E)^4 \ \frac{-8.0 \ i \pi}{2 (ig_E)^2} \ e^{-\pi/g_E} \; ,\\
        &\mathcal{I}^{(2R)}_1 = \frac{q^2 m^4}{1024 \pi^4} \ (-ig_E)^6 \ \frac{0.1  \pi}{2 (ig_E)^3} \ e^{-\pi/g_E} \; .
    \end{align}
\end{subequations}
The comparison for the 1-PI case is presented in fig.\,\ref{fig:L2I-electric-field}, revealing that the first instanton contribution computed from the weak-field expansion via the Borel dispersion relation provides a sufficiently accurate approximation to the behaviour of the imaginary part of the full 1-PI action. 

Interestingly, this ceases to be the case for the 1-PR action. The kink observed in the imaginary part of the 1-PR effective action, which signifies a change in the function's sign, is not effectively replicated by the first instanton contribution constructed from the weak-field expansions. However, such behaviour may possibly be accommodated by sub-leading corrections to the large-order behaviour of $a_n^{(2R)}$.  Let us model the sub-leading corrections to be of the form
\begin{multline}
    a_n^{(2R)} \sim 0.1 (-1)^n \ \frac{\Gamma(2n+3)}{\pi^{2n+3}} + (-1)^n \ C \ \frac{\Gamma(2n+k)}{\pi^{2n+k}} \\ ~\ ,~~ n \to \infty \; ,
\end{multline}
assuming $k<3$. This sub-leading correction to the large order behaviour of the weak-field coefficients will consequently lead to a correction to the instanton of the form (taking $k_{\mathcal{P}}=k-1$ in eq.\,(\ref{eq:temp0811232}))
\begin{equation}\label{eq:1-PR:instanton-contribution}
    \mathcal{I}^{(2R)}_1 = \frac{q^2 m^4}{1024 \pi^4} \ g_E^6 \left[-\frac{0.1i \pi e^{-\pi/g_E}}{2 g_E^3} + C \frac{i\pi e^{-\pi/g_E}}{2 g_E^k} \right] \; .
\end{equation}

This instanton correction, while suppressed in the regime of weak fields, can rapidly overtake the leading instanton contribution for sufficiently strong fields and can, in fact, reverse the sign of $\mathcal{I}^{(2R)}_{1}$. 

We note purely empirically that a fitted correction to first instanton contribution with $k=1.5$ and $C = 0.027$ (plotted in fig.\,\ref{fig:L2R-electric-field-imaginary-part}) is a much better approximation to the non-perturbative behaviour of $\mathcal{L}^{(2)}_{\text{1-PR}}$ than the uncorrected instanton contribution.

\begin{figure*}[t]
    \begin{subfigure}{0.55\linewidth}
        \includegraphics[width=0.9\linewidth]{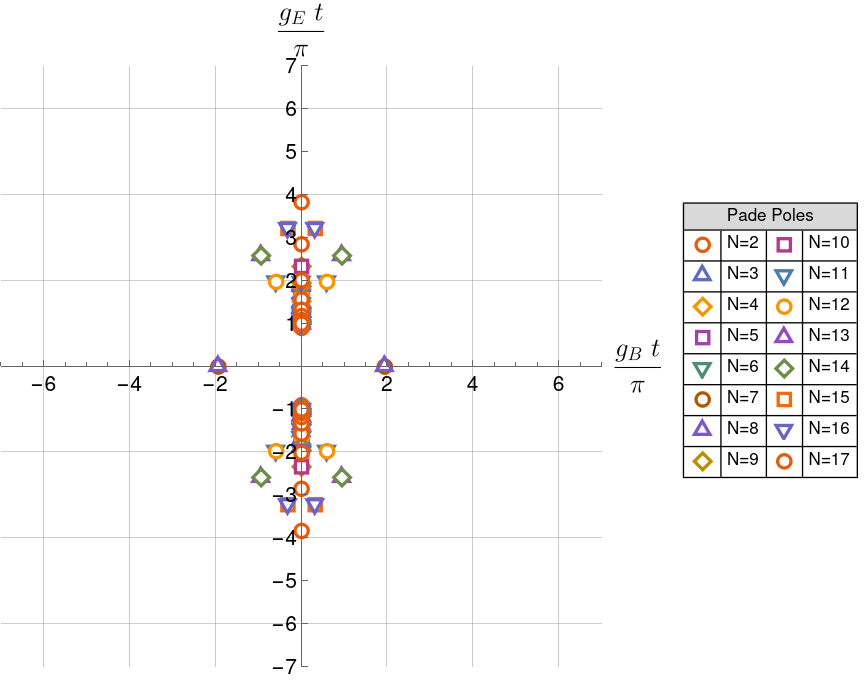}
        \caption{}
        \label{sfig:L2R-poles}
    \end{subfigure}
    \begin{subfigure}{0.35\linewidth}
        \includegraphics[width=\linewidth]{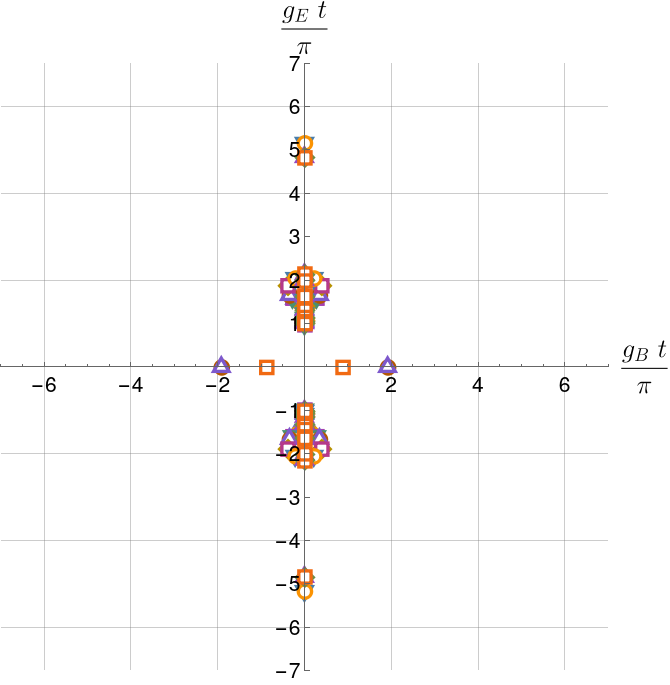}
        \caption{}
        \label{sfig:L2I-poles}
    \end{subfigure}
    \caption{Poles of the Pad\'{e} approximants of the Borel sum for the 1-PI (fig. \ref{sfig:L2R-poles}) and 1-PR (fig. \ref{sfig:L2I-poles}) contributions. The presence of branch cuts is suggestive in both singularity structures.}
    \label{fig:two-loop-poles}
\end{figure*}

One of the consequences of the Borel dispersion relations (as is shown at the end of Appendix \ref{app:borel-dispersion-relations}) is that a fractional fit for $k$ implies the presence of algebraic singularities that manifest as branch cuts in the pole structure of the 2-loop sQED 1-PR Borel transforms. In the case of spinor QED 1-PI reconstructions, a similar feature has also been commented on recently\,\cite{dunne_higher-loop_2021}. Indeed, the presence of branch cuts are suggested when the singularities of the Pad\'e approximants of $\widehat{\mathcal{S}}_{N^*}^{(2I)}$ and $\widehat{\mathcal{S}}_{N^*}^{(2R)}$ are plotted on the complex plane in fig.\,\ref{fig:two-loop-poles}. The presence of accumulated simple poles of the Pad\'e-approximants, seen near integer multiples of $\pi$, is an attempt at mimicking the properties of the multi-valued function despite being single-valued themselves. In addition to poles on the imaginary axis, spurious poles surround the poles accumulated on the imaginary axis. As in the case of the 1-loop sQED EHL analysis, these spurious poles are pushed to infinity in the limit $N\to \infty$. 

The weak field expansions in eqs.\,(\ref{eq:1-PI-weak-field-expansion}) and (\ref{eq:1-PR-weak-field-expansion}) indicate that for weak fields, the 1-PI contributions dominate over their 1-PR counterparts, and one could potentially ignore the latter contribution in this limit. However, it has recently been pointed out\,\cite{Gies:2016yaa} that the 1-PR diagrams can compete with the contributions of the 1-PI diagrams in the strong-field regime. This is indeed also seen when we compare the values attained by the two contributions in figs.\,\ref{fig:two-loop-1PI-diff-mp} and \ref{fig:two-loop-1PR-diff-mp}. We observe that for $g_E \sim \mathcal{O}(10^3)$, the 1-PI and 1-PR contributions to the EHL for a magnetic field background are of the same order. Already at the order of $g_E \sim 10^3$, we realize that the 1-PR contributions cannot be ignored. 

Another curious property of the 1-PR diagrams in sQED is the kink in its imaginary contribution to the 2-loop EHL, as demonstrated in fig.\,\ref{fig:L2R-electric-field-imaginary-part}. This kink signifies a change of sign for $g_E\sim 2$. Also, the instanton contribution in eq.\,(\ref{eq:1-PR:instanton-contribution}) suggests that while the value of the imaginary part of $\mathcal{L}^{(2)}_{\text{1-PR}}$ is negative for small value of $g_E$, the imaginary part changes sign for a sufficiently large $g_E$. Physically, this implies that while 1-PR diagrams hinder the vacuum decay rate in the weak field regime by contributing negatively to the imaginary part of the EHL, they start facilitating the pair production rate after a critical electric field value.  

\section{Summary and conclusions}  \label{sec:conclusions}
The program of resurgent extrapolations, resummation methods, approximants and trans-series\,\cite{ecalle1985fonctions,10.1093/oso/9780198535850.002.0003, Sternin:1996,Costin:2008,DORIGONI2019167914,Aniceto:2018bis,baker1975essentials,baker1996pade} have been subjects of intense research in recent years. At the same time, the study of quantum field theories in strong-field regimes has profound theoretical and experimental relevance, especially with the advent of new high-intensity LASER facilities and observation of astrophysical environments where such large fields are routinely present\,\cite{Altarelli:2019zea, Hattori:2023egw, Fedotov_2023,Ruffini:2009hg,Kaspi:2017fwg,Kim:2019joy,Grasso:2000wj,Durrer:2013pga}. The present work lies at the interface of these two broad topics.

In this work, we investigated resurgence and Padé-Borel resummation in the theory of scalar quantum electrodynamics by looking at the resurgent extrapolation of its weak-field expansions. There have already been a few pioneering works analysing spinor QED\,\cite{florio, dunne_higher-loop_2021,non-linear-trident-resummation,Torgrimsson:2022ndq,Dunne:2022esi,Torgrimsson:2021wcj,Torgrimsson:2021zob} in the context of resurgent extrapolations. Remarkably, in contrast to spinor quantum electrodynamics, scalar electrodynamics has not been studied in much detail in these settings so far. There have only been a few preliminary comments in the existing literature\,\cite{Huet:2017ydx}. Apart from the fact that scalar quantum electrodynamics is a relatively well-understood theory, thereby helping us compare the Padé-Borel reconstructed functions to the full theory relatively easily, it also may be a prototype for potential models in the dark matter sector, like millicharged particles\,\cite{HOLDOM1986196,doi:10.1146/annurev.nucl.012809.104433}.

We looked carefully at the 1-loop and 2-loop contributions to the scalar quantum electrodynamics Euler-Heisenberg Lagrangian in the Schwinger proper time framework and their corresponding weak-field expansions. In these contexts, we re-derived a few well-known expressions in the literature in forms amenable to resurgent analysis while also deriving for the first time closed-form expressions for others, such as the one-particle reducible contribution at two loops.

We then investigated in some detail Padé-Borel resummation methods to reconstruct the 1-loop and 2-loop Euler-Heisenberg Lagrangian of scalar quantum electrodynamics using only a finite number of terms from their weak-field expansions. We tried different schemes of Pad\'{e}-Borel reconstructions in each case and commented on the optimal prescription. 

Among the distinctive aspects of our analysis was the inclusion of one-particle reducible diagrams in the 2-loop Euler-Heisenberg Lagrangian of scalar quantum electrodynamics. These contributions were thought to be irrelevant until recently. We compared our reconstructions of the 1-loop and 2-loop Euler-Heisenberg Lagrangian with the derived analytic closed forms and results from numerical computations. Remarkably, it is found that even a few terms from the weak-field expansions may be sufficient to reconstruct the full function resurgently. Although there are simple relations between the 1-loop EHL of QED and sQED, and the 1-loop sQED EHL and 2-loop 1-PR sQED EHL, the resurgent structures do not seem to follow these simple relations and have non-trivial differences in the pole structures.

As in the case of spinor quantum electrodynamics, we found that the scalar quantum electrodynamics Padé-Borel reconstructed Euler-Heisenberg Lagrangian seems to lose its meromorphic properties at the 2-loop order. This occurs both for the one-particle irreducible and one-particle reducible parts. For both spinors and scalars, this is a distinction between the 1-loop and 2-loop Borel transform counterparts. The Borel transform of the former only has simple poles in both cases. We analyzed the trans-series structure of both these at the order of the first instanton contribution.

Through analytic continuation to the electric field case, we studied the appearance of instabilities in the theory, manifested as scalar particle Schwinger pair production. By analyzing the large-order behaviour and through the application of Borel dispersion relations, we were able to derive the imaginary parts explicitly. Additionally, our findings revealed that the one-particle reducible contributions to the vacuum decay rate have the surprising behaviour of suppressing the decay rate in the weak field limit for $qE/m^2 \lesssim 2$ and enhancing the decay rate in the strong-field regime. It would be interesting to explore the implications of this behaviour in physical settings.

Apart from the germane discussions and expressions in Secs.\,\ref{sec:ehl}\;and\;\ref{sec:pade-borel-analysis-of-ehl}, some of our main results are encapsulated in figs.\,\ref{fig:L1-diff-mp},\,\ref{fig:L1-Asy-vs-Pade},\,\ref{fig:L1-reconstruction-electric},\,\ref{fig:two-loop-1PI-diff-mp},\,\ref{fig:two-loop-1PR-diff-mp},\,\ref{fig:L2I-electric-field},\,\ref{fig:L2R-electric-real-part},\;and\;\ref{fig:L2R-electric-field-imaginary-part}. In summary, in this work, we studied in relative detail and successfully checked the efficacy of Padé-Borel reconstructed Euler-Heisenberg Lagrangians in scalar quantum electrodynamics.

There may be a few avenues for future explorations. Gaining a better understanding of the singularity and trans-series structures from higher-loop orders in scalar electrodynamics may be a fertile direction for future research. Additionally, investigating the resurgent properties of these trans-series structures using mathematical tools such as alien derivatives and Stoke's automorphisms might provide more insight and clarity in the exposition of these subjects. An intriguing question is also what the contribution of the $(\phi^*\phi)^2$ term, required for renormalizability of sQED, will be to the Euler-Heisenberg Lagrangian. Another interesting avenue will be exploring finite temperature and matter density effects, which are of physical relevance in many scenarios of interest. 

\section*{Aknowledgements}
DG acknowledges support from an INSPIRE fellowship from the Department of Science and Technology, Government of India.
\clearpage

\appendix 

\section{Closed form for integral expressions}\label{app:closed-form-of-integral}
It is well known that integrals like the ones in eqs. (\ref{eq:1-loop:integral-form}) and (\ref{eq:two-loop-1PR-integral}) can be expressed in the closed form in terms of Hurwitz-Zeta functions $\zeta_H(s,a)$\,\cite{DUNNE_2005,titchmarsh1986theory}. For completeness, we will derive these forms in this appendix, in our conventions and notations, following the ideas developed in\,\cite{Karbstein_2015,Karbstein_2013}. 

Let us start with the integral in eq.\,(\ref{eq:1-loop:integral-form}), which we will denote by $I_1$:
\begin{equation}\label{app-eq:I1}
    I_1 = \int_0^\infty \frac{ds}{s^2} \ e^{-m^2s/qB} \left[\frac{1}{\sinh s} - \frac{1}{s} + \frac{s}{6} \right] \; .
\end{equation}
There are some identities that we will use extensively in the coming derivations (see the identities in\,\cite[p.~251]{apostol1998introduction})
\begin{subequations}
    \begin{align}
        &\int_0^\infty ds \ s^{\nu-1} e^{-\beta s} = \beta^{-\nu} \ \Gamma(\nu) \; , \\
        &\int_0^\infty ds \  s^{\nu -1} \frac{1}{\sinh s} \ e^{-\beta s} = 2^{1-\nu} \ \Gamma(\nu) \ \zeta_H\left(\nu,\frac{1+\beta}{2} \right) \; . 
    \end{align}
\end{subequations}
These identities are only valid for $\text{Re} \ \beta >0$ and $\text{Re} \ \nu >0$, and hence cannot be applied to eq.\,(\ref{app-eq:I1}) directly. To perform an appropriate analytic continuation, we introduce the parameter $\epsilon$ in the integral:
\begin{equation}\label{app-eq:I1-epsilon-form}
    I_1 = \lim_{\epsilon \to 0} \int_0^{\infty} \frac{ds}{s^{2-\epsilon}} \ e^{-m^2s/qB} \left[\frac{1}{\sinh s} - \frac{1}{s} + \frac{s}{6} \right] \; .
\end{equation}
The individual integrals in eq.\,(\ref{app-eq:I1-epsilon-form}) can now be performed and, in the leading and first subleading order, be written as
\begin{multline}
        \int_0^{\infty} \frac{ds}{s^{2-\epsilon}} \ \frac{1}{\sinh s} \ e^{-m^2 s/qB} =  \\ - \frac{4}{\epsilon} \ \zeta_H \left( -1, \frac{1+m^2/qB}{2}\right)  \\  + \zeta_H \left(-1,\frac{1+m^2/qB}{2} \right) (4\log 2 -4 + 4 \gamma) \\ -  4  \zeta_H' \left(-1,\frac{1+m^2/qB}{2} \right) \; ,
\end{multline}
where the derivative of the Hurwitz-zeta function is with respect to the first argument and
\begin{multline}
    \int_0^{\infty} \frac{ds}{s^{3-\epsilon}} \ e^{-m^2s/qB} = \\ \frac{1}{2} \left( \frac{m^2}{qB} \right)^2 \left[\frac{1}{\epsilon} + \frac{3}{2} - \gamma - \log \left(\frac{m^2}{qB} \right) \right] \; ,
\end{multline}
\begin{equation}
    \int_0^{\infty} \frac{ds}{s^{1-\epsilon}} \ e^{-m^2s/qB} = \frac{1}{\epsilon} - \log \left(\frac{m^2}{qB} \right) - \gamma \; .
\end{equation}
Using the property of the zeta function,
\begin{equation}
    \zeta_H \left(-1,\frac{1+x}{2} \right) = \frac{1}{24} - \frac{x^2}{8} \; ,
\end{equation}
we see that the divergent parts in $I_1$ cancel out, and the finite part that remains is 
\begin{multline}
    I_1 = -4 \left[\zeta_H'\left(-1,\frac{1+m^2/qB}{2} \right) + \right. \\  \left. \zeta_H\left(-1,\frac{1+m^2/qB}{2} \right) \left( \log \left(\frac{m^2}{2qB} \right)+1\right) + \right. \\  \left. \frac{3}{16} \left(\frac{m^2}{qB}\right)^2\right] \; .
\end{multline}
Now, we will try to perform similar manipulations on the integral in eq.\,(\ref{eq:two-loop-1PR-integral}). In this case, the following manipulation using integration by parts will come in handy:
\begin{equation}
\begin{split}
    &\int_0^{\infty} \frac{ds}{s^3} \ e^{-m^2s/qB} \left[\frac{s}{\sinh s}- \frac{s^2}{\sinh s \tanh s} + \frac{s^2}{3} \right] \\ &= \int_0^{\infty} \frac{ds}{s^2} \ e^{-m^2s/qB} \  \dv{s} \left[\frac{s}{\sinh s} + \frac{s^2}{6} -1\right]  \\
    &= \frac{m^2}{qB} \int_0^{\infty} \frac{ds}{s^2} \  e^{-m^2s/qB} \left[\frac{s}{\sinh s} + \frac{s^2}{6} -1 \right]+\\  &2\int_0^{\infty} \frac{ds}{s^3} \ e^{-m^2s/qB} \left[ \frac{s}{\sinh s}+ \frac{s^2}{6}-1 \right] \; .
\end{split}
\end{equation}
We can now calculate the closed form $I_2$ using the same techniques we used above to get that
\begin{multline}
    I_2 = -4 \zeta_H \left(-1,\frac{1+m^2/qB}{2} \right)  \\ - 8 \zeta_H' \left(-1,\frac{1+m^2/qB}{2} \right)   \\ + 2 \left(\frac{m^2}{qB}\right)^2 \zeta_H' \left(0,\frac{1+m^2/qB}{2} \right) \\  -\frac{1}{3} \log \left(\frac{m^2}{2qB} \right) \; .
\end{multline}
\section{Derivation of 1-PI effective action in Schwinger proper time} \label{app_sec:1PI-effective-action} 
To simplify eq.\,(\ref{eq:two-loop-1-PI:intial_expression}), we need to evaluate the second functional derivative of $\overline{W}^{(1)}$ with respect to the background gauge field. We already know the first functional derivative from eq.\,(\ref{eq:1-loop-noether-current-limit-form}),
\begin{equation}
    \frac{\delta \overline{W}^{(1)}}{\delta A_\mu (x)} = \pdv{\mathcal{L}_\phi}{A_\mu} \ (x) = -2iq \lim_{x \to y} D^\mu_x G_A(x-y) \; .
\end{equation}
From this expression, it is apparent that we will need the functional derivative of $G_A$ with respect to $A_\mu(x)$ to proceed further in the calculations. To do this, we start with the Green's function equation,
\begin{equation} \label{eq:temp010623}
        (D_z^2 + m^2)G_A(z-y) = -i \delta^4(z-y) \; ,
\end{equation}
where $D^{\mu}_z = \partial^\mu_z +iqA^\mu(z)$ is the covariant derivative. Now, we multiply by the free scalar propagator $G_F(x-z)$ on both sides and integrate over the $z$ space, expand the covariant derivative, and use the identity 
\begin{multline}
    G_F(x-z) \partial_z^2 G_A(z-y) = G_A(z-y) \partial_z^2 G_F(x-z) \\ - \partial_z(G_A(z-y) \partial_z G_F(x-z)) \\ + \partial_z(G_F(x-z) \partial_z G_A(z-y)) 
\end{multline}
to get the relation
\begin{multline} \label{app-eq:temp0308231}
        -i G_A(x-y) - q^2 \int d^4z \ A^2(z) G_A(z-y) G_F(x-z)  \\ + 2iq \int d^4z \ A(z) G_F(x-z) \partial_z G_A(z-y) \\ +  iq \int d^4z \ \partial A(z) G_F(x-z) G_A(z-y) = \\  -i G^F(x-y) \; .
\end{multline}
This equation, in the operator notation, can be written as
\begin{equation}
        \widehat{G}_A = \widehat{G}_F + iq^2 \widehat{G}_F\widehat{A}^2 \widehat{G}_A - 2iq \widehat{G}_F \widehat{A} \widehat{p} \widehat{G}_A + q \widehat{G}_F \partial \widehat{A} \widehat{G}_A  \; ,
\end{equation}
where $\widehat{p}^\mu$ is the momentum operator, $\widehat{A}^\mu=A^\mu(\widehat{x})$, the operator $\widehat{G}_A$ is defined as $\bra{x} \widehat{G}_A\ket{y} = G_A(x-y)$, and $\widehat{G}_F$ is defined in a similar manner. It will be useful to keep in mind that this equation can be rearranged to give
\begin{equation} \label{app-eq:temp0308232}
    \widehat{G}_A = (1-iq^2 \widehat{G}_F\widehat{A}^2 + 2iq \widehat{G}_F\widehat{A}\widehat{p} - q \widehat{G}_F \partial \widehat{A}  )^{-1} \widehat{G}_F \; .
\end{equation}
Now, we will take a functional derivative on both sides of eq.\,(\ref{app-eq:temp0308231}) to get that
\begin{multline}
        \frac{\delta G_A(x-y)}{\delta A_\mu(w)} = \\ iq^2 \int d^4z \ \Bigg[ 2G_F(x-z)A^\mu(z) \delta^4(z-w) G_A(z-y)  \\  + G_F(x-z)A^2(z) \frac{\delta G_A(z-y)}{\delta A_\mu(w)}  \Bigg] \\ + 2q \int d^4z \ \Bigg[G_F(x-z) \delta^4(z-w) \partial_z G_A(z-y)  \\  + G_F(x-z) A(z) \partial_z \frac{\delta G_A(z-y)}{\delta A_\mu(w)} \Bigg] \\ + q \int d^4z \ \Bigg[G_F(x-z)  \partial^\mu_z(\delta^4(z-w)) G_A(z-y) \\  + G_F(x-z) \partial A(z)  \frac{\delta G_A(z-y)}{\delta A_\mu(w)} \Bigg] \; .
    \end{multline}
A couple of simplifications can be done to arrive at the final answer--
\begin{multline}
    \partial^\mu_z ( \delta^4(z-w)) G_A(x-z) G_A(z-y) = \\ - \delta^4(z-w) \partial^\mu_z ( G_A(x-z) G_A(z-y))  \; ,
\end{multline}
and 
\begin{equation}
    \begin{split}
        \partial^\mu_z \bra{x} \widehat{G}_A \ket{z} &= ( \partial^\mu_z \bra{z} \widehat{G}_A^\dagger \ket{x})^* \; , \\
        &= (-i \bra{z} \widehat{p}^\mu \widehat{G}_A^\dagger \ket{x} )^* \; ,\\
        &= i \bra{x} \widehat{G}_A \widehat{p}^\mu \ket{z} \; .
    \end{split}
\end{equation}
After this simplification, we will use eq.\,(\ref{app-eq:temp0308232}) to arrive at the first functional derivative of $G_A$,
\begin{multline}
        \frac{\delta \widehat{G}_A}{\delta A_\mu (w)} = -iq \bigg[ \bigg(\widehat{G}_A \ket{w} \bra{w} \widehat{\Pi}^\mu \widehat{G}_A \bigg) + \\ \bigg( \widehat{G}_A \widehat{\Pi}^\mu \ket{w} \bra{w} \widehat{G}_A \bigg) \bigg] \; ,
\end{multline}
where $\widehat{\Pi}^\mu = \widehat{p}^\mu - \widehat{A}^\mu$. Substituting this in eq.\,(\ref{eq:two-loop-1-PI:intial_expression}), we have that
\begin{multline}
    W^{(2)}_{\text{1-PI}}[A] = -iq^2 \int d^4x d^4y  \bigg[4i \delta^4(x-y) \bra{x} \widehat{G}_A \ket{y} \\ - \bra{x} \widehat{\Pi}^\mu \widehat{G}_A \ket{y} \bra{y} \widehat{\Pi}_\mu \widehat{G}_A \ket{x} - \\ \bra{x} \widehat{\Pi}^\mu \widehat{G}_A \widehat{\Pi}_\mu \ket{y} \bra{y}\widehat{G}_A \ket{x}  \bigg] D_+(x-y) \; .
\end{multline}
This expression can be computed by substituting the values of the photon and dressed electron propagator in proper time form\,\cite{Schwinger_1951}--
\begin{equation}
    D_+(x-y) = \frac{i}{16 \pi^2} \int_0^{\infty} \frac{ds}{s^2} \ \exp \left[\frac{-i(x-y)^2}{4s}  - \varepsilon s \right] \; ,
\end{equation}
and
\begin{multline}
    G_A(x-y) =  - \frac{i}{16 \pi^2} \Phi(x,y) \\ \int_0^\infty \frac{ds}{s^2} \ e^{-s \epsilon} \exp \left[-\frac{i}{4} z^{\mu} \beta_{\mu \nu}(s) z^{\nu} - L(s) -ism^2 \right] \; , 
\end{multline}
where $\Phi(x,y)$ is the gauge-dependent phase the final answer is independent of, $z=x-y$, and the matrix $\beta$ and scalar $L(s)$ is given by
\begin{subequations}
\begin{align}
    &\beta = q F \coth(qFs) \; ,\\
    &L(s) = \frac{1}{2} \text{Tr} \log [(qFs)^{-1}\sinh (qFs)] \; .
\end{align}
\end{subequations}
($F$ without the tensor indices denotes the electromagnetic matrix that follows the conventional rules of matrix multiplication). Now, we can easily compute that
\begin{equation}\label{app-eq:temp1410231}
\begin{split}
    &\bra{x} \widehat{\Pi}^\mu G_A \ket{y} = (i\partial_\mu - q A_\mu) G_A(x-y) \; , \\
            &= -\frac{i}{16 \pi^2} \Phi(x,y) \int_0^{\infty} \frac{ds}{s^2} \ e^{-s \epsilon} \frac{1}{2} \bigg(qF^{\mu \nu}z_\mu \\ &+ \beta^{\mu\nu}(s)z_\nu \bigg)  \exp \left[-\frac{i}{4} z \beta(s) z - L(s) -ism^2 \right] \; ,
\end{split}
\end{equation}
\begin{equation}\label{app-eq:temp1410232}
\begin{split}
    &\bra{x} G_A \widehat{\Pi}^\mu \ket{y} = \bigg( \bra{y} \widehat{\Pi}_\mu (\widehat{G}_A)^* \ket{x} \bigg)^* \; , \\
            &= -\frac{i}{16 \pi^2} \Phi(x,y) \int_0^{\infty} \frac{ds}{s^2} \ e^{-s \epsilon} \frac{1}{2} \bigg(-qFz+ \beta(s)z \bigg) \\ &\exp \left[-\frac{i}{4} z \beta(s) z - L(s) -ism^2 \right] \; .
\end{split}
\end{equation}
To find $\bra{x}\widehat{\Pi}^\mu G_A \widehat{\Pi}_{\mu} \ket{y}$, we can use a relation that is apparent from eqs.\,(\ref{app-eq:temp1410231}) and (\ref{app-eq:temp1410232})---
\begin{equation}
    \widehat{G}_A \widehat{\Pi}_\mu = \widehat{\Pi}_\mu \widehat{G}_A - qFz \widehat{G}_A \; .
\end{equation}
Multiplying both sides by $\widehat{\Pi}^\mu$, taking an expectation value, and using Green's equation for the evolution of $G_A$, we have the simplified relation
\begin{multline}
    \bra{x}\widehat{\Pi}^\mu \widehat{G}_A \widehat{\Pi}_\mu \ket{y} = i \delta^4(x-y) + m^2 G_A(x-y) \\ - \frac{q^2}{2} (z FFz) G_A(x-y) \; .
\end{multline}
The multiplicative term $e^{-L(s)}$ can be simplified as
\begin{equation}\label{app-eq:temp1410233}
    \begin{split}
            &\exp \left[- \frac{1}{2} \text{Tr}  \log \left(\frac{\sinh(qFs)}{qFs} \right) \right] \\ &= \text{det} \left[ \exp \left(- \frac{1}{2} \log \left(\frac{\sinh(qFs)}{qFs} \right)\right) \right]  \; ,\\
            &= \text{det} \left[ \frac{qFs}{\sinh(qFs)} \right]^{1/2} \; ,\\
            &= \frac{q^2s^2  \eta \epsilon}{\sin(q \eta s) \sinh(q \epsilon s)} \; ,
        \end{split}
\end{equation}
where $\eta$ and $\epsilon$ are the eigenvalues of $F^{\mu \nu}$\,\cite{Schwinger_1951}. All the terms that contain the delta function factor $\delta^4(x-y)$ are divergent and, in fact, renormalize the mass of the 1-loop EHL. Finally, there are two types of terms left in the action:
\begin{equation}
    R_1 =  - m^2  \int d^4z G_A(z) D(z) G_A(-z) \; ,
\end{equation}
and 
\begin{multline}
    R_2 = \\ \int d^4z \ \left( \frac{3q^2}{4}zFFz + \frac{1}{4} z \beta \beta z \right) G_A(z) D(z) G_A(-z) \; .
\end{multline}
Using eq.\,(\ref{app-eq:temp1410233}) and straightforward Gaussian integrations, we arrive at the unrenormalized result\,\cite{v_i_ritus_connection_1977}
\begin{multline}
    \mathcal{L}^{(2)}_{\text{1-PI}} =  \frac{i q^2}{256 \pi^4} \int_0^\infty ds \int_0^{\infty} ds' \\ \frac{(q\eta)^2 (q \epsilon)^2 e^{-i(s+s')(m^2 - i \epsilon)}}{\sin(q\eta s)\sin(q \eta s')\sinh(q \epsilon s)\sinh(q \epsilon s')} \\ \left( m^2 I_0 + \frac{iI}{2} \right) \; ,
\end{multline}
where 
\begin{subequations}
\begin{align}
    &I = \frac{(q - p) \log(b/a)}{(b-a)^2} - \frac{\frac{q}{b}  - \frac{p}{a}}{b-a} \; ,\\
    &I_0 = \frac{1}{b-a} \log \left(\frac{b}{a} \right) \; ,
\end{align}
\end{subequations}
with the definitions
\begin{subequations}
    \begin{align}
        &a = q \eta \cot(q \eta s) + q \eta \cot(q \eta s') \; , \\
        &b = q \epsilon \coth(q \epsilon s) + q \epsilon \coth(q \epsilon s') \; ,\\
        &p = 2 (q \eta)^2 ( \cot(q \eta s)  \cot(q \eta s')+3) \; , \\
        &q = 2 (q \epsilon)^2 (\coth(q \epsilon s)  \coth(q \epsilon s') - 3) \; .
    \end{align}
\end{subequations}
As in the case of the 1-loop EHL, the divergences in this integral are due to the vacuum energy and a term that is proportional to the free Lagrangian. These divergences can be removed by expanding the integrand in the limit $\eta\to 0$, $\epsilon \to 0$ and subtracting the resultant divergences. After a lengthy procedure of renormalization, contour rotation, and the change of variables
\begin{subequations} 
        \begin{align}
            t &= s + s' \; ,\\
            u &= \frac{s'}{s+s'} \; ,
        \end{align}
\end{subequations}
we get the final result stated in eq.\,(\ref{eq:two-loop-1-PI-integral-form}).

\section{Borel Dispersion relations}\label{app:borel-dispersion-relations} 
\begin{figure*}[th!]
\centering
\begin{subfigure}{0.45\linewidth}
    \includegraphics[width=\linewidth]{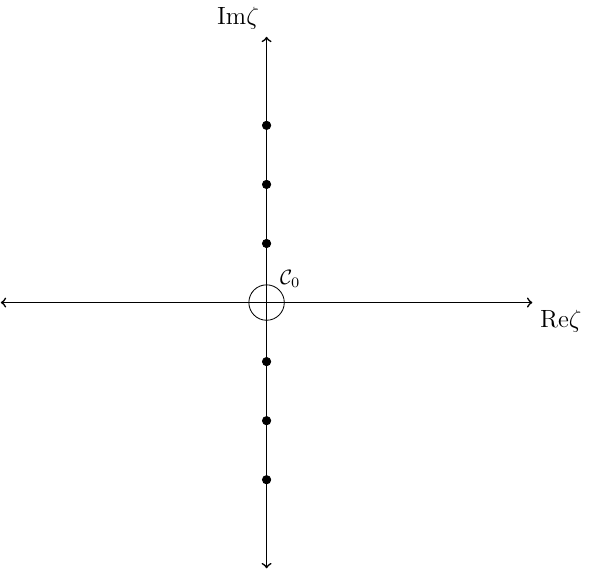}
    \caption{}
    \label{fig:hankel-undeformed}
\end{subfigure}
\begin{subfigure}{0.45\linewidth}
\includegraphics[width=\linewidth]{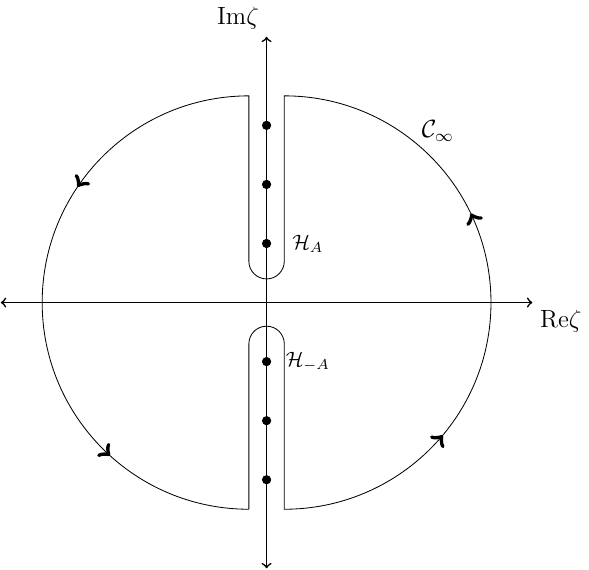}
\caption{}
\label{fig:modified-hankel-contour}
\end{subfigure}
    \caption{The contours used for deriving the Borel dispersion relation in eq.\,(\ref{app-eq:borel-dispersion-relation}). The Cauchy contour integral in fig.\,\ref{fig:hankel-undeformed} is deformed to surround the poles of the Borel sum $\widehat{\mathcal{S}}(\zeta)$ in fig. \ref{fig:modified-hankel-contour}. The poles of $\widehat{\mathcal{S}}$ are denoted by filled circles on the imaginary axis.}
    \label{fig:hankel-contour}
\end{figure*}
The Borel dispersion relations establish a link between the asymptotic behavior of weak-field coefficients and the non-perturbative dynamics of the Borel sum observed in the strong-field regime. Consider the formal series in eq.\,(\ref{eq:def-formal-divergent-series}). For our purposes, we will state the modified result of the Borel dispersion relation for weak-field coefficients that have large-order asymptotics of the form
\begin{equation}\label{eq:temp0711231}
    a_n \sim  \ \sum_{\mathcal{P}=1}^{\infty} C_{\mathcal{P}} \ (iA_{\mathcal{P}})^{-2n-k_{\mathcal{P}}-1} \ \Gamma(2n+k_{\mathcal{P}}+1) \ , \ n \to \infty \; .
\end{equation}
The corresponding Borel transform,
\begin{equation}
    \widehat{\mathcal{S}}_{p,N^*}(z) = \sum_{n=0}^{N^*} \frac{a_n}{(2n+p)!} z^{2n+p} \; ,
\end{equation}
will have singularities at $\pm iA_{\mathcal{P}}$ in the limit $N^* \to \infty$. The singularity structure of $\hat{S}_{p,N^*}$ can be, for example, algebraic, such that
\begin{equation}
        \widehat{\mathcal{S}}_p (\pm iA_{\mathcal{P}} \pm i\xi) \sim (\pm 1)^{p} \ \widetilde{C}_{\mathcal{P}} \ (-\xi)^{-b} \ , \ \xi \to 0 \; ,
    \end{equation}
or can have a simple pole such that
\begin{equation}
        \widehat{\mathcal{S}}_p (\pm iA_{\mathcal{P}} \pm i\xi) \sim -(\pm 1)^{p} \ \frac{\widetilde{C}_{\mathcal{P}}}{i\xi}  \ , \ \xi \to 0 \;,
\end{equation}
where $\widehat{\mathcal{S}}_p = \lim_{N^* \to \infty} \widehat{\mathcal{S}}_{p,N^*}$. To find out the nature of the singularity, we consider a contour integral of the function $\widehat{\mathcal{S}}_p$ around the origin, along the contour shown in fig.\,\ref{fig:hankel-undeformed}. By the Cauchy integral theorem, we know that
\begin{equation}
    \frac{a_n}{\Gamma(2n+p+1)} = \frac{1}{2\pi i} \oint_{\mathcal{C}_0} d\zeta \ \frac{\widehat{\mathcal{S}}_p(\zeta)}{\zeta^{2n+p+1}} \; .
\end{equation}
Now, this contour near the origin can be deformed in the way described in fig.\,\ref{fig:hankel-contour} to obtain the so-called ``Hankel contour" (see, for example,\,\cite{krantz2012handbook,marino_2015}). This contour contains two parts-- the circular part $\mathcal{C}_{\infty}$, the integration over which vanishes for a very large radius, and the Hankel contours $\mathcal{H}_A$ and $\mathcal{H}_{-A}$ around the singularities of $\widehat{\mathcal{S}}$. The integral around the Hankel contours $\mathcal{H}_A$ and $\mathcal{H}_{-A}$ have the same contribution because of the symmetry between the poles $iA_{\mathcal{P}}$ and $-iA_{\mathcal{P}}$. Using all of these facts, we can now easily do the contour integral over $\mathcal{H}_A$ using the singularity structure of $\widehat{\mathcal{S}}$. For instance, when $\widehat{\mathcal{S}}$ only has simple poles, one gets the result
\begin{equation}
     \frac{a_n}{\Gamma(2n+p+1)} \sim  2 \sum_{\mathcal{P}=1}^{\infty}  \widetilde{C}_{\mathcal{P}} (iA_{\mathcal{P}})^{-2n-p-1}  \ , \  n \to \infty \; .
\end{equation}
Comparing this with the large-order behaviour that we started with in eq.\,(\ref{eq:temp0711231}), we get that
\begin{equation}\label{eq:temp0811231}
    p = k_{\mathcal{P}} \; ,
\end{equation}
and 
\begin{equation}
    \widetilde{C_{\mathcal{P}}}  = \frac{C_{\mathcal{P}}}{2} \; . 
\end{equation}
Now, we can also use the singularity structure of $\widehat{\mathcal{S}}$ to find out its contribution to the imaginary part of $\text{B}_p[\mathcal{S}]$. To do this, consider the Borel transform in eq.\,(\ref{eq:def-borel-transform}) with $z = -ig_E$ with the variable change $t \to t/z$,
\begin{multline}
    \text{B}_p[\mathcal{S}](g_E) = \frac{1}{(-ig_E)^{p+1}} \times \\ \lim_{N^* \to \infty} \int_0^{-i \infty} dt \ e^{-it/g_E} \ \widehat{\mathcal{S}}_{p,N^*}(\zeta) \;.
\end{multline}
The imaginary part of this integral can be calculated by performing contour integral around the singularities of $\widehat{\mathcal{S}}$ on the negative imaginary axis to get that,
\begin{equation}
    2 \ \text{Im} [\text{B}_p[\mathcal{S}]](g_E)  = \frac{2 \pi i}{(ig_E)^{p+1}} \sum_{\mathcal{P}=0}^{\infty} \widetilde{C}_{\mathcal{P}} e^{-A_{\mathcal{P}}/g_E} \; .
\end{equation}
Using these relations, we get a neat relation that gives the imaginary part of $\text{B}[\mathcal{S}]$ in terms of $C_{\mathcal{P}}$ and $k_{\mathcal{P}}$--
\begin{equation}\label{app-eq:borel-dispersion-relation}
    \text{Im} [\text{B}_p[\mathcal{S}]](g_E) = \frac{\pi i}{2} \sum_{\mathcal{P}=1}^{\infty} \frac{C_{\mathcal{P}}}{(ig_E)^{k_{\mathcal{P}}+1}} e^{-A_{\mathcal{P}}/g_E} \; .
\end{equation}
This is the Borel dispersion relation. It can be proved that the same relation is obtained for the case of algebraic and logarithmic singularity as well. It will be useful to note that when this procedure is adopted in the case of an algebraic singularity, we get a relation between $b$, $k_{\mathcal{P}}$, and $p$ akin to eq.\,(\ref{eq:temp0811231})--
\begin{equation}
    k_{\mathcal{P}} = p+b-1 \; .
\end{equation}
Note the significance of this relation-- it says that if $k_{\mathcal{P}}$ is fractional, $b$ must also be fractional. Therefore, fractional values of $k_{\mathcal{P}}$ lead to algebraic singularities. 

\section{Coefficient values for the weak field expansion of $\mathcal{L}^{(2)}_{\text{1-PI}}$ and $\mathcal{L}^{(2)}_{\text{1-PR}}$ for sQED}\label{app:2-loop-coefficient-values}
\begin{table*}[b!]
\centering
    \begin{tabular}{c|c|c}
         $n$ & $a_n^{(2I)}$ & $a_n^{(2R)}$ \\
    \hline
      0 & $\frac{275}{648}$ & $\frac{49}{8100}$\\
      1 & $-\frac{5159}{16200}$  & $-\frac{31}{2700}$ \\
      2 & $\frac{751673}{1058400}$  & $\frac{58433}{1587600}$ \\
      3 &  $-\frac{931061}{291600}$ &  $-\frac{5161351}{26195400}$ \\
      4 & $\frac{139252117469}{5762988000}$ & $\frac{629915089}{378378000}$\\
      5 & $-\frac{1749526192619}{6324318000}$ & $-\frac{35555913023}{1702701000}$\\
      6 & $\frac{60143228485253}{13471920000}$ & $\frac{2806953752810761}{7641722088000}$\\
      7 & $-\frac{342211558219703077}{3542980240800}$ & $-\frac{552956876118767}{64017954000}$\\
      8 & $\frac{111676851457699685948647}{41379925106520000}$ & $\frac{15229529901011542321}{58256338140000}$\\
      9 & $-\frac{7924897558024929739223}{83765030580000}$& $-\frac{650336486466192684637}{65776701790800}$\\
      10 & $\frac{121800315777228315482407336783}{29954709995530320000}$& $\frac{428964436921204922260687}{939667168440000}$\\
      11 & $-\frac{31118443383101537277412172791}{147997579029300000}$& $-\frac{3012135397426829997023747}{119265448302000}$\\
      12 & $\frac{1372366357724439120548020044517}{106558256901096000}$ & $\frac{6785551823434972770541748948323}{4115850620902020000}$\\
      13 & $-\frac{12360815173481596296265109135631547}{13404026888392140000}$& $-\frac{8462680145742528698979498502657}{67508660977758000}$\\
      14 & $\frac{101430749192220049807898294140516375455307}{1328991702086778649920000}$& $\frac{3876135337081567976230866367314231534199}{352968763887563791968000}$\\
      15 & $-\frac{38745242595653082092152082819293043866847}{5358837508414430040000}$& $-\frac{3885536678316440235723614259967162597697}{3539125306359262620000}$
    \end{tabular}
\caption{A tabulation of the first fifteen coefficients of the weak-field expansion of the 1-PI and 1-PR contributions to the 2-loop EHL of sQED.}
\label{table:1-PI-coefficients }
\end{table*}
While the weak-field coefficients of the 1-loop EHL have a closed form and can be computed up to arbitrary order, the weak-field expansions of the 2-loop EHL don't have any closed form. They are computationally challenging to compute, even numerically. We list 15 of the numerically computed weak-field coefficients of both 1-PI and 1-PR contributions of the 2-loop EHL for the case of sQED in table\,\ref{table:1-PI-coefficients }.
\clearpage
\bibliography{SQEDResurgence}
\end{document}